\documentclass[12pt]{article}
\usepackage[english]{babel}
\usepackage[latin1]{inputenc}

\usepackage[intlimits]{amsmath}
\usepackage{amssymb}

\usepackage{url}
\usepackage{graphicx}
\usepackage{slashed}
\usepackage{color}

\numberwithin{equation}{section}

\newcommand{\GeV}{\mbox{GeV}}
\newcommand{\TeV}{\mbox{TeV}}
\newcommand{\cm}{\mbox{cm}}

\def\lsim{\mathrel{\raise.3ex\hbox{$<$\kern-.75em\lower1ex\hbox{$\sim$}}}}
\def\gsim{\mathrel{\raise.3ex\hbox{$>$\kern-.75em\lower1ex\hbox{$\sim$}}}}

\begin{document}


\title{\vspace{-2cm} 
{\normalsize
\flushright DESY 11-257 \\
\vspace{-.4cm}\flushright TUM-HEP 823/11\\}
\vspace{0.6cm} 
\bf Dark matter annihilations into two light fermions and one gauge boson:\\ general analysis and antiproton constraints\\[8mm]}

\author{Mathias Garny$^1$, Alejandro Ibarra$^2$, Stefan Vogl$^2$\\[2mm]
{\normalsize\it  $^1$ Deutsches Elektronen-Synchrotron DESY, Hamburg}\\[-0.05cm]
{\normalsize\it  Notkestra\ss{}e 85, 22603 Hamburg, Germany}\\[1.5mm]
{\normalsize\it $^2$Physik-Department T30d, Technische Universit\"at M\"unchen,}\\[-0.05cm]
{\it\normalsize James-Franck-Stra\ss{}e, 85748 Garching, Germany}
}

\maketitle

\begin{abstract}
\noindent
We study in this paper the scenario where the dark matter is constituted by Majorana particles which couple to a light Standard Model fermion and an extra scalar via a Yukawa coupling. In this scenario, the annihilation rate into the light fermions  with the mediation of the scalar particle is strongly suppressed by the mass of the fermion. Nevertheless, the helicity suppression is lifted by the associated emission of a gauge boson, yielding annihilation rates which could be large enough to allow the indirect detection of the dark matter particles. We perform a general analysis of this scenario, calculating the annihilation cross section of the processes $\chi\chi\rightarrow f\bar f V$ when the dark matter particle is a $SU(2)_L$ singlet or doublet, $f$ is a lepton or a quark, and $V$ is a photon, a weak gauge boson or a gluon. We point out that the annihilation rate is particularly enhanced when the dark matter particle is degenerate in mass to the intermediate scalar particle, which is a scenario barely constrained by collider searches of exotic charged or colored particles. Lastly, we derive upper limits on the relevant cross sections from the non-observation of an excess in the cosmic antiproton-to-proton ratio measured by PAMELA. 
\end{abstract}

\newpage

\section{Introduction}

Among the various proposals to characterize the dark matter in our Universe, the scenario where the dark matter is constituted by weakly interacting Majorana particles stands as the most promising one. In this scenario, thermal scatterings of Standard Model particles in the early Universe can produce a relic density of dark matter particles which is of the correct order of magnitude, when the interaction strength is of the order of the weak interaction strength and the dark matter mass is about 1 TeV. Furthermore, this scenario has the appealing feature that the dark matter particle might be directly detected in underground detectors, indirectly detected in cosmic ray detectors, gamma-ray and neutrino telescopes, and directly produced at the LHC.

In this paper we will focus on the possibility of indirectly detecting Majorana dark matter particles via their self-annihilation in the Milky Way dark matter halo. Concretely we will study the annihilation process into two fermions and one gauge boson which, under some circumstances, can have a non-negligible or even a larger cross section than the lowest order annihilation process into two fermions. More specifically, the $s$-wave contribution to the thermally averaged cross section for the $2\rightarrow 2$ process is helicity suppressed by the mass of the final fermion, while the $p$-wave contribution is suppressed by the small velocity of the dark matter particles in the Milky Way halo. In contrast, for the $2\rightarrow 3$ process the $s$-wave contribution is no longer suppressed, due to the associated emission of a vector in the final state. As a result, the $2\rightarrow 3$ processes can even have a larger cross section than the $2\rightarrow 2$ processes as the lifting of the helicity suppression can compensate the suppression due to the additional coupling $\alpha_{em}/\pi$, provided the mediating scalar particles are not too heavy.

Dark matter annihilations into two fermions and one photon with the mediation of a heavy scalar particle were first studied in the framework of the Minimal Supersymmetric Standard Model in neutralino annihilations \cite{Bergstrom:1989jr,Flores:1989ru}, and explored in a number of subsequent papers~\cite{Bringmann:2007nk}. It was also pointed out that this process not only could have a sizable cross-section but also produces a gamma-ray with a very peculiar spectral shape which, if detected in gamma-ray telescopes, could be unequivocally identified as being originated in dark matter annihilations~\cite{Scott:2009jn,Bringmann:2011ye}. If the mediating scalar particle is electrically charged it must necessarily carry hypercharge and therefore the annihilation process must also produce weak gauge bosons and, in turn, antiprotons. This process has been studied in \cite{Ciafaloni:2011sa,Garny:2011cj} employing a toy model where the dark matter particle is a singlet under the Standard Model gauge group and the mediating particle is a $SU(2)_L$ doublet. The constraints on the annihilation cross section in this model from the non-observation of an excess in the cosmic antiproton-to-proton fraction measured by PAMELA were derived in \cite{Garny:2011cj}. Lastly, annihilations into quarks, so that the emission of a gluon is allowed in the final state were discussed in \cite{Flores:1989ru,Drees:1993bh}.

In this paper we aim to extend this analysis, considering various toy models where the dark matter particle is a $SU(2)_L$ singlet or doublet, which couples to the left-handed or right-handed leptons or quarks of the first generation. For each case, we will calculate the cross sections for the different $2\rightarrow 3$ processes and we will calculate the constraints on the cross sections from the PAMELA measurements of the antiproton-to-proton fraction~\cite{Adriani:2010rc}. 

The paper is organized as follows. In section \ref{sec:leptons}, we discuss in detail the cross sections for the various electromagnetic and electroweak internal bremsstrahlung processes occurring for $SU(2)_L$ singlet as well as doublet dark matter particle coupling to leptons. The corresponding cases, when assuming a coupling to quarks, are discussed in section \ref{sec:quarks}. In section \ref{sec:pbar} we present the constraints on the cross sections from the PAMELA measurements of the antiproton-to-proton fraction, and translate them into upper limits on an astrophysical boost factor for the various toy models. Finally, we conclude in section \ref{sec:conclusions}. Our full analytical results for the cross sections of all $2\to 3$ processes considered in this work can be found in the Appendix.

\section{Dark matter coupling to leptons}\label{sec:leptons}

We consider an extension of the Standard Model by one Majorana fermion, $\chi$, which we assume to constitute the dominant component of dark matter in the Universe, and one scalar particle,  $\eta$, which mediates the annihilation process into
light fermions. The Lagrangian is
\begin{align}
{\cal L}={\cal L}_{\rm SM}+{\cal L}_{\chi}+{\cal L}_\eta+
{\cal L}^{\rm fermion}_{\rm int}+{\cal L}^{\rm scalar}_{\rm int}\;.
\end{align}
Here, ${\cal L}_{\rm SM}$ is the Standard Model Lagrangian which includes a potential for the Higgs doublet $\Phi$, $V=m_1^2 \Phi^\dagger \Phi +\frac{1}{2}\lambda_1 (\Phi^\dagger \Phi)^2$. On the other hand ${\cal L}_\chi$ and ${\cal L}_\eta$ are the parts of the Lagrangian involving just the Majorana fermion $\chi$ and the scalar particle $\eta$, respectively, and which are given by
\begin{align}
\begin{split}
{\cal L}_\chi&=\frac12 \bar \chi^c i\slashed{\partial} \chi
-\frac{1}{2}m_\chi \bar \chi^c\chi\;, \\
{\cal L}_\eta&=(D_\mu \eta)^\dagger  (D^\mu \eta)-m_2^2 \eta^\dagger\eta-
\frac{1}{2}\lambda_2(\eta^\dagger \eta)^2\;,
\end{split}
\end{align}
where $D_\mu$ denotes the covariant derivative. Lastly, ${\cal L}^{\rm fermion}_{\rm int}$ and ${\cal L}^{\rm scalar}_{\rm int}$ denote the fermionic and scalar interaction terms of the new particles to the leptons and to the Higgs doublet. These terms depend on the details of the model and will be discussed case by case below.

The quantum numbers of the relevant Standard Model particles under the gauge group $SU(3)_C\times SU(2)_L\times U(1)_Y$ are: $e_R\equiv(1,1,-1)$, $L_e\equiv(1,2,-\frac{1}{2})$,  $\Phi\equiv(1,2,\frac{1}{2})$. On the other hand, the quantum numbers of the dark matter particle and the scalar $\eta$ are constrained in our setup by the requirement that the dark matter particle is colorless and electrically neutral, and by the requirement that a Yukawa coupling to the leptons (either left-handed or right-handed) is invariant under the Standard Model gauge group. We will assume in the following that the dark matter particle only couples to the first generation of leptons, which can be ensured by postulating that the extra scalar particle $\eta$ carries electron lepton number $L_e=-1$, while the dark matter particle does not carry lepton number. Lastly, in order to guarantee the stability of the dark matter particle, we impose a $Z_2$ discrete symmetry under which $\chi$ and $\eta$ are odd while the Standard Model particles are even.

In the scenarios of interest for this paper, the intermediate scalar is electrically charged and could possibly lead to experimental signatures in collider experiments. Precise measurements of the invisible decay width of the $Z$ boson at LEP set the upper bound $\Delta \Gamma_{\rm inv}<2.0$ MeV~\cite{:2005ema}, which rules out the existence of exotic charged scalar particles with mass below 40 GeV~\cite{Nakamura:2010zzi}. Furthermore, the OPAL collaboration searched for an excess with respect to the Standard Model expectations of dilepton events with missing energy induced by the production of exotic scalar charged particles which decay into an electron and an invisible particle (in the framework of supersymmetry, the production of selectrons which decay into an electron and the lightest neutralino). The non-observation of an excess in a sample of 680 pb$^{-1}$ of $e^+e^-$ collisions at center-of-mass energy between 192 GeV and 209 GeV, leads to the lower bound $m_{\eta}\geq 97.5$ GeV, assuming $m_{\eta}-m_{\rm DM}>11$ GeV~\cite{Abbiendi:2003ji}. A similar search was undertaken by the L3 collaboration using a sample of 450 pb$^{-1}$ collisions at $\sqrt{s}=183-209$ GeV, resulting in the lower bound $m_{\eta}\geq 94.4$ GeV assuming $m_{\eta}-m_{\rm DM}>10$ GeV~\cite{Achard:2003ge}, by the ALEPH collaboration using a sample of 207 pb$^{-1}$ collisions at $\sqrt{s}=204-209$ GeV, resulting in $m_{\eta}\geq 95$ GeV assuming $m_{\eta}-m_{\rm DM}>15$ GeV~\cite{Heister:2001nk} and by the DELPHI collaboration using a sample of 609 pb$^{-1}$ collisions at $\sqrt{s}=192-208.8$ GeV, resulting in $m_{\eta}\geq 94$ GeV assuming $m_{\eta}-m_{\rm DM}>15$ GeV, and $m_{\eta}\geq 98$ GeV assuming $m_{\eta}-m_{\rm DM}>5$ GeV and $m_{\rm DM}<60$ GeV~\cite{Abdallah:2003xe}. For smaller mass splittings the detection efficiency is significantly reduced and the lower bounds derived by the LEP experiments can be avoided.

In the remainder of this section we present a classification of models,  according to the charge of the dark matter particle under $SU(2)_L$.

\subsection{$SU(2)_L$ singlet dark matter}

When the dark matter particle is a $SU(2)_L$ singlet, its hypercharge must be zero in order to render an electrically neutral particle, hence the gauge quantum numbers must be $\chi=(1,1,0)$. On the other hand, the gauge quantum numbers of the scalar $\eta$ depend on whether the dark matter particle couples to the right-handed electron singlet or to the left-handed electron doublet.

If the dark matter has a Yukawa coupling to the right-handed electron singlet and a scalar field $\eta$, then gauge invariance requires $\eta=(1,1,1)$. With these quantum numbers the only interaction terms in the Lagrangian are:
\begin{align}
  \begin{split}
    {\cal L}^{\rm fermion}_{\rm int}&= -f \bar \chi e_R \eta+{\rm h.c.}\;,  \\
    {\cal L}^{\rm scalar}_{\rm int}&= -\lambda_3(\Phi^\dagger \Phi)
    (\eta^\dagger \eta)\;.
  \end{split}
\label{eq:singlet-eR}
\end{align}
In the Minimal Supersymmetric Standard Model (MSSM) this possibility
is realized if $\chi$ is the bino and $\eta$ is 
the right-handed selectron, $\tilde e_R$.

On the other hand, if the dark matter only couples to the left-handed electron doublet then $\eta=(1,2,-\frac{1}{2})$. The interaction terms are then:
\begin{align}
  \begin{split}
    {\cal L}^{\rm fermion}_{\rm int}&= -f \bar \chi(L_e i\sigma_2 \eta)+{\rm h.c.}=
    -f \bar \chi (\nu_{eL} \eta^0-e_L \eta^+)+{\rm h.c.}\;,  \\
    {\cal L}^{\rm scalar}_{\rm int}&= -\lambda_3(\Phi^\dagger \Phi)
    (\eta^\dagger \eta)
    -\lambda_4(\Phi^\dagger \eta)(\eta^\dagger \Phi)\;.
  \end{split}
\label{eq:singlet-L}
\end{align}
In the MSSM, this possibility is realized if $\chi$ is the bino and $\eta$ the left-handed selectron doublet, $\tilde L_e$.

After the electroweak symmetry breaking, the mass of the electrically charged scalar $\eta^\pm$ is given by $m_{\eta^\pm}^2=m_2^2+\lambda_3v_{EW}^2$. If the dark matter couples to the left-handed electron doublet, there exists also a neutral scalar with mass given by $m_{\eta^0}^2=m_2^2+(\lambda_3+\lambda_4)v_{EW}^2$. We will assume that $m_{\chi}<m_{\eta^i}$, such that the Majorana fermion $\chi$ is stable and can constitute the dark matter. The interactions lead to a thermal production of $\chi$ in the Early Universe that is compatible with the WMAP value if the coupling is of order one $f\sim\mathcal{O}(1)$ and the masses lie between the weak and the TeV scale~\cite{Cao:2009yy,Garny:2011cj}.

The annihilation of dark matter in the Milky Way today can proceed via the $2\to 2$ annihilation channels $\chi\chi\to e\bar e$, as well as $\chi\chi\to \nu\bar \nu$. However, the cross-sections $\sigma v_{2\to 2}=a+bv^2$ are highly suppressed because the $s$-wave contribution $a\propto m_e^2$ is helicity suppressed, while the $p$-wave contribution $bv^2$ is suppressed by the dark matter velocity $v\sim10^{-3}c$ in the Milky Way halo. The helicity suppression can be lifted by emitting an additional spin-1 particle in the final state \cite{Bergstrom:1989jr,Flores:1989ru,Bringmann:2007nk,Ciafaloni:2011sa,Garny:2011cj}.  The lifting of the helicity suppression can compensate the suppression due to the additional coupling $\alpha_{em}/\pi$ provided the mediating scalar particles are not too heavy, typically $m_{\eta^i}\lesssim 5m_{\chi}$~\cite{Garny:2011cj}. Therefore, the dominant annihilation processes are $2\to 3$ channels, like $\chi\chi\to \gamma e\bar e$, $\chi\chi\to Z e\bar e$  or $\chi\chi\to W e\bar \nu$. 

Note that one could similarly consider a coupling of the dark matter particle to the leptons of the second or third generation. The cross-sections for the $2\to 3$ processes and the constraints from the antiproton flux that will be discussed later are independent of the lepton flavor to a good accuracy. Due to the larger masses of the $\mu$ and $\tau$ leptons, the helicity suppression of the $2\to 2$ annihilation cross-sections is less pronounced than for a coupling to electrons, while the velocity suppressed contributions to the $2\to 2$ annihilation cross-sections are flavor-independent.

The annihilation mode $\chi\chi\to \gamma e\bar e$ leads to a gamma ray signal with a pronounced peak at the dark matter mass, that is potentially observable by the Fermi-LAT (see~\cite{Bringmann:2012vr} for a recent discussion) and by current and future IACTs~\cite{Scott:2009jn,Bringmann:2011ye}. On the other hand, the annihilation channels involving weak gauge bosons yield a primary contribution to the cosmic flux of antiprotons. In the following, we will analyze the relative strength of these channels, and in Section 4 we will derive upper limits on the cross-section from the PAMELA measurement of the antiproton to proton ratio~\cite{Adriani:2010rc}.

\subsubsection*{Coupling to right-handed electrons}

Let us first consider the possibility that the dark matter particle $\chi$ couples to right-handed electrons. Since the mediating particle $\eta$ carries hypercharge and electric charge, the $2\to 3$ annihilation channels $\gamma e\bar e$ and $Z e\bar e$ are possible \cite{Ciafaloni:2011sa,Garny:2011cj}. The branching ratios compared to the $2\to 2$ annihilation cross section are shown in the upper panel of Fig.~\ref{fig:crossSection1}. For dark matter masses $m_{\rm DM}\gg \frac{M_Z}{2}$, far above the $Z$-threshold, the ratio of the cross sections for the electromagnetic and the electroweak bremsstrahlung processes approach constant values given by the ratio of the respective coupling constants:
\begin{equation}
\begin{array}{ccccccc}
  \sigma v(\chi\chi\to Z e\bar e) & : & \sigma v(\chi\chi\to \gamma e\bar e) & = & \tan^2(\theta_W) & \simeq & 0.30 \;.
\end{array}
\end{equation}
The general formulas for the double differential cross sections and for arbitrary dark matter masses are given in the Appendix. In order to obtain a gauge-invariant result it is important to take into account the diagrams for which the gauge boson is emitted off the internal line and off the final state particles (later on, when considering doublet dark matter, also contributions from initial state radiation have to be included). The corresponding Feynman diagrams are also shown in the Appendix. In the following, we will refer to all these processes as internal bremsstrahlung (IB). As noted before, these comprise also the contributions from final state radiation in general. Note that the contributions from soft and collinear emission, which are in principle logarithmically enhanced, are typically negligible in this context because they are helicity suppressed, like the $2\to 2$ processes. Instead, the dominant contribution arises from the diagrams where the gauge boson is emitted either off the internal line, or from a final state particle with an off-shell intermediate state (see Ref.~\cite{Garny:2011cj} for a detailed discussion).

The dependence of the cross sections on the mass $m_\eta$ is shown in the upper right panel of Fig.~\ref{fig:crossSection1}. It is apparent from the figure that the branching ratio of the $2\to 3$ processes is largest when $m_\eta$ is close to $m_{\rm DM}$. Furthermore, for $\mu\equiv m_\eta^2/m_{\rm DM}^2\gg 1$, the cross section of the $2\to 3$ processes fall of as $1/\mu^4$, while the $2\to 2$ cross sections scale like $1/\mu^2$. The former remain dominant as long as the dark matter mass and the mass of the mediating particle are of comparable size, roughly $m_\eta\lesssim 5m_{\rm DM}$. The qualitative properties discussed here are common also to most other cases considered below. However, there are some quantitative and also qualitative differences which we will stress in the following.

\subsubsection*{Coupling to left-handed electrons}

\begin{figure}
\hspace*{-1.5cm}
\begin{tabular}{ll}
 \includegraphics[width=0.55\textwidth]{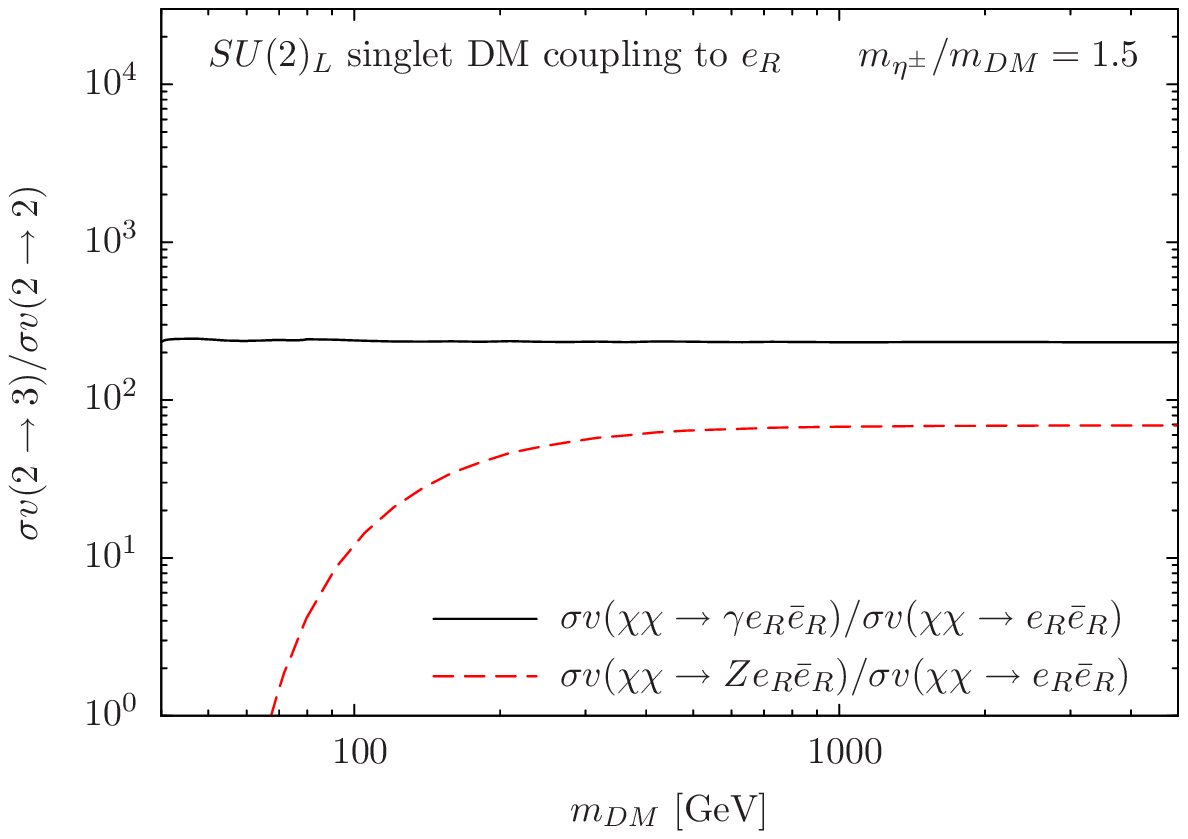}
 &\hspace*{-0.8cm} \includegraphics[width=0.55\textwidth]{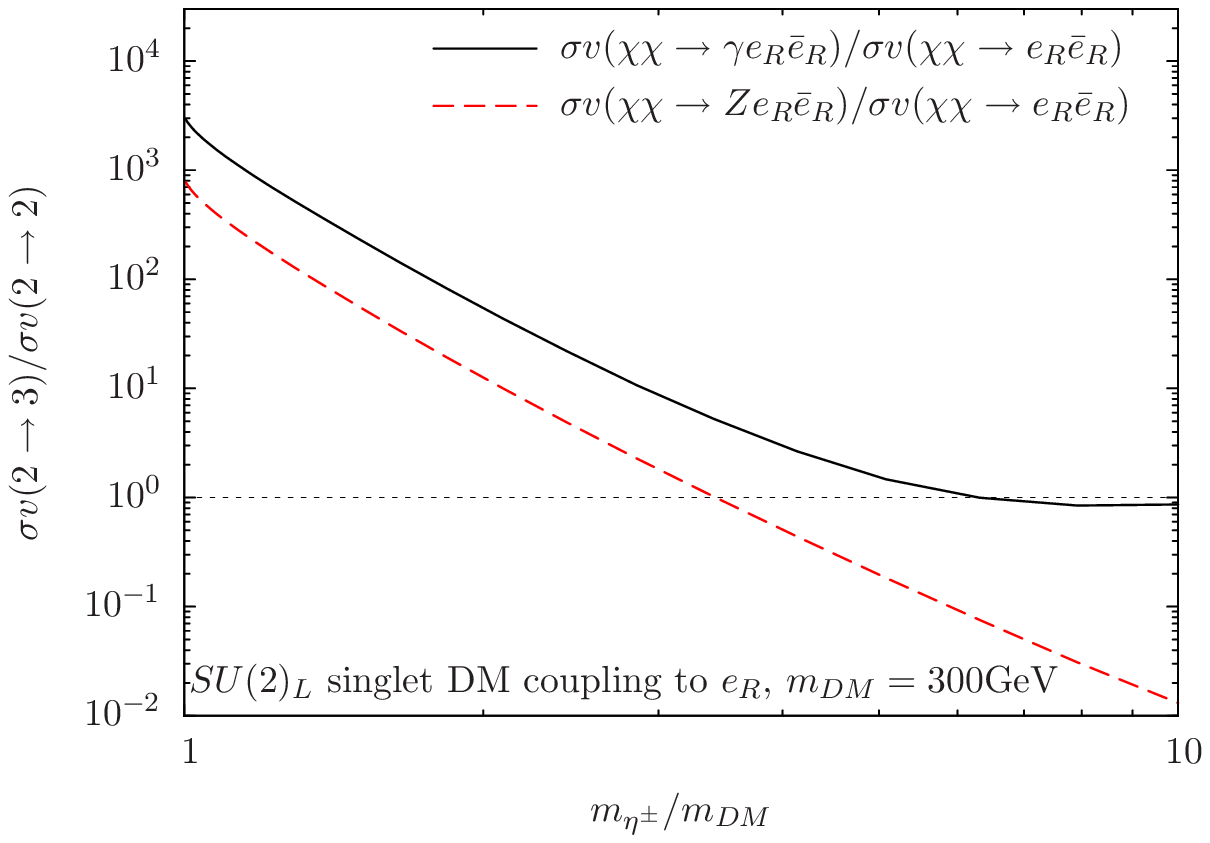} \\
 \includegraphics[width=0.55\textwidth]{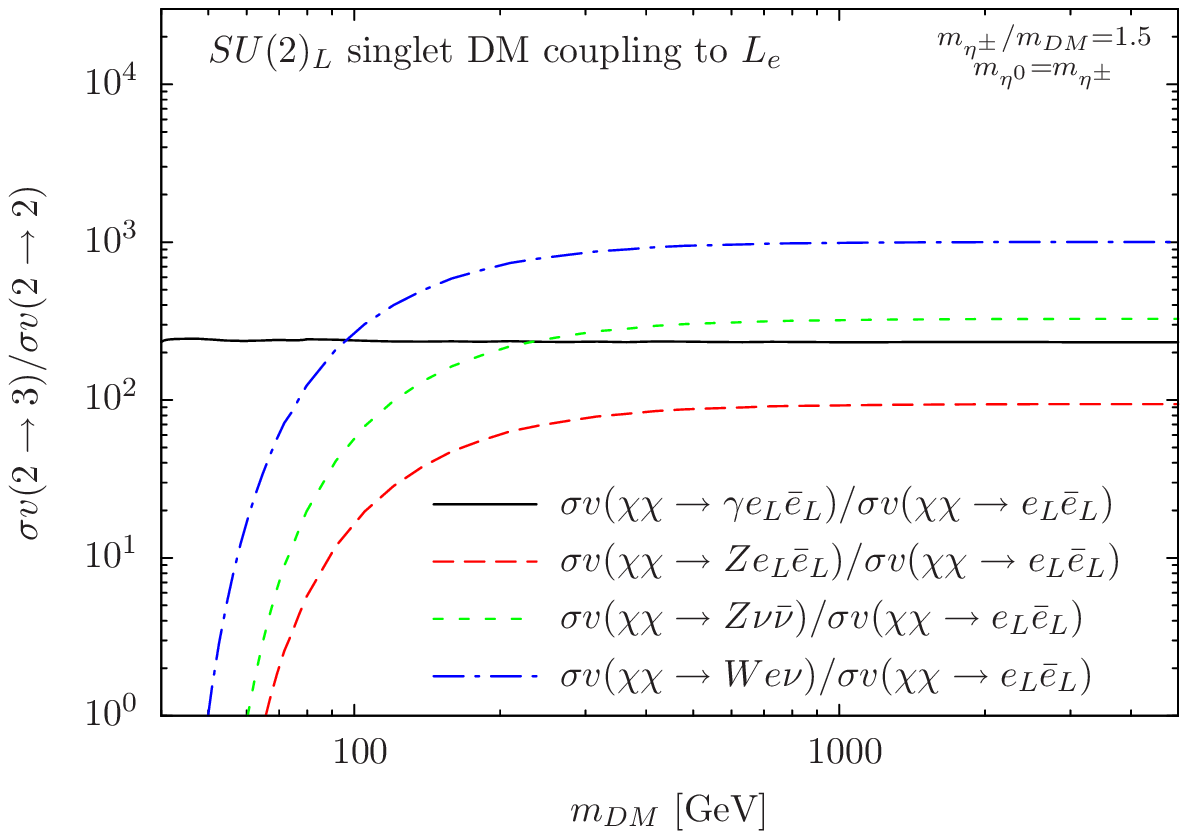}
 &\hspace*{-0.8cm} \includegraphics[width=0.55\textwidth]{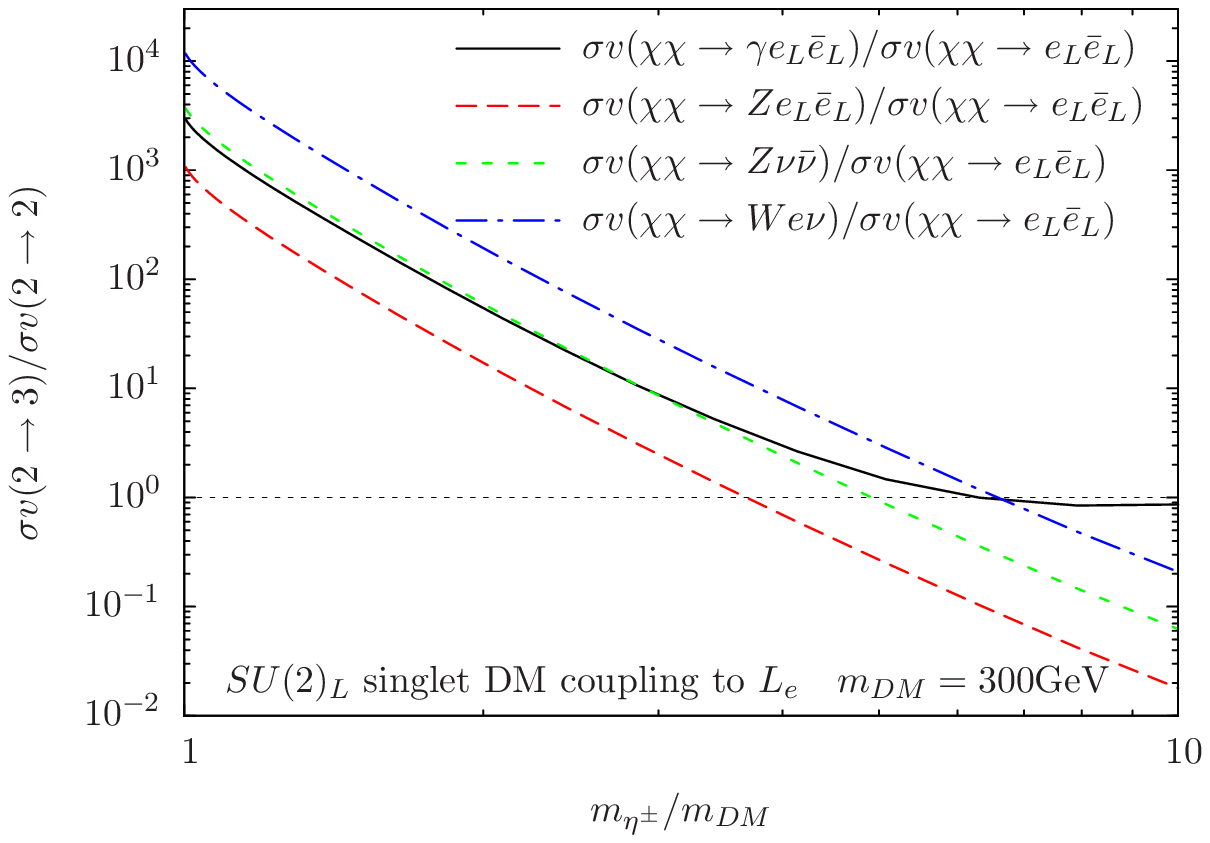} \\
\end{tabular}
 \caption{\label{fig:crossSection1} Ratio of three-body and two-body annihilation cross-sections, for electromagnetic IB, $\sigma v(\chi\chi\to \gamma e\bar e)/\sigma v(\chi\chi\to e\bar e)$, and for the electroweak IB channels $\chi\chi\to Ze\bar e, \chi\chi\to Z\nu\bar\nu$ and $\chi\chi\to We\nu$. The latter denotes the sum of $W^- \bar e\nu$ and $W^+ e\bar \nu$. The top and bottom rows show the case of $SU(2)_L$ singlet dark matter coupling to the right-handed electron, and to the left-handed electron doublet, respectively. The left column shows the dependence on the dark matter mass for fixed ratio $m_{\eta^\pm}/m_{\rm DM}=1.5$, while the right column shows the dependence on the mass of the mediating scalar particle $\eta$ for $m_{\rm DM}=300$GeV. For the relative dark matter velocity we use $v=10^{-3}c$.}
\end{figure}

If the dark matter particle $\chi$ couples to the left-handed electron doublet, it can give rise to annihilations into final states involving $\gamma, Z$ or $W$ bosons. Note that this case has been discussed in detail in Ref.~\cite{Garny:2011cj}. We will briefly review it here for completeness. The branching ratios compared to the $2\to 2$ annihilation cross section are shown in the lower part of Fig.~\ref{fig:crossSection1}. For $m_{\rm DM}\gg \frac{M_Z}{2}$, and assuming that $m_{\eta^0}=m_{\eta^\pm}$, the asymptotic values are again given by the ratios of the appropriate couplings:
\begin{equation}
\begin{array}{ccccccc}
  \sigma v(\chi\chi\to Z e\bar e) & : & \sigma v(\chi\chi\to \gamma e\bar e) & = & \cot^2(2\theta_W) & = & 0.41\,, \\
  \sigma v(\chi\chi\to Z \nu\bar \nu) & : & \sigma v(\chi\chi\to \gamma e\bar e)
  & = & \frac{1}{\sin^{2}(2\theta_W)} & = & 1.41\,, \\
  \sigma v(\chi\chi\to W e\nu) & : & \sigma v(\chi\chi\to \gamma e\bar e)
  & = & \frac{1}{\sin^{2}(\theta_W)} & = & 4.32 \,.
\label{eq:ratios-lepto}
\end{array}
\end{equation}
Here $\sigma v(\chi\chi\to W e\nu)\equiv \sigma v(\chi\chi\to W^- \bar e\nu) + \sigma v(\chi\chi\to W^+ e\bar \nu)$.

Generically, one expects a non-zero mass splitting of the neutral and charged components of $\eta$ induced by the breaking of the electroweak symmetry, $m_{\eta^0}^2-m_{\eta^\pm}^2=\lambda_4 v_{EW}^2$. Compared to the degenerate limit, the branching ratios get modified due to two effects. First, the masses in the t-channel propagators of the mediating particles corresponding to the charged and the neutral component of $\eta$ differ from each other. Second, the mass splitting opens up a new channel, namely the annihilation into longitudinally polarized $W$-bosons. In the limit $m_{\eta^i} \gg m_{\rm DM}\gg \frac{M_Z}{2}$, the branching ratios are approximately given by
\begin{eqnarray}
  \frac{\sigma v(\chi\chi\to Z \nu\bar \nu) }{ \sigma v(\chi\chi\to \gamma e\bar e) }
  & \simeq & \frac{1}{\sin^{2}(2\theta_W)} \frac{\mu_\pm^4}{\mu_0^4}\;, \nonumber\\
  \frac{ \sigma v(\chi\chi\to W e\nu) }{ \sigma v(\chi\chi\to \gamma e\bar e) }
  & \simeq & \frac{1}{\sin^{2}(\theta_W)}  \frac{\mu_\pm^4}{\mu^4}\left[ 1 + \frac{5}{8}\frac{m_{\rm DM}^2}{ M_W^2}(\mu_\pm-\mu_0)^2  \right]\;.
\end{eqnarray}
where $\mu_i=m_{\eta^i}^2/m_{\rm DM}^2$ and $\mu=(\mu_\pm+\mu_0)/2$. The ratio $\sigma v(\chi\chi\to Z e\bar e) / \sigma v(\chi\chi\to \gamma e\bar e)$, in contrast, is not affected by the mass splitting. The emission of longitudinal $W$ bosons also leads to a spectrum that is harder compared to the case $m_{\eta^0}=m_{\eta^\pm}$~\cite{Garny:2011cj}. Analytical expressions for the double differential cross sections, from which the spectra can be easily obtained, are given in the Appendix.

\subsection{$SU(2)_L$ doublet dark matter}\label{doublet}

In this case the dark matter doublet must have one electrically neutral component, which is achieved by postulating that the hypercharge is $\pm 1/2$. This matter content, however, leads to gauge anomalies which can be canceled by introducing another $SU(2)_L$ doublet with opposite hypercharge. Then, the minimal model with $SU(2)_L$ doublet fermionic dark matter must contain the new fermions $\chi_1\equiv (1,2,-\frac{1}{2})$, $\chi_2\equiv (1,2,\frac{1}{2})$, both charged under the $Z_2$ discrete symmetry. In the case of the MSSM these two particles can be identified with the two higgsinos.

Under these assumptions, the only gauge invariant and $Z_2$ invariant fermionic mass term in the Lagrangian is $M\bar\chi_1^ci\sigma_2\chi_2+h.c.$, which generates identical tree level masses for $\chi_1^0$, $\chi_1^-$, $\chi_2^0$, $\chi_2^+$. Quantum corrections induced by the Standard Model gauge bosons generate a mass splitting between the charged and the neutral component of the multiplet $(m_{\chi^\pm}-m_{\chi^0})_{rad}\simeq 0.34$GeV \cite{Cirelli:2005uq}, inducing the decay of the former into the latter. Therefore, this toy model predicts the existence of two stable particles, candidates of dark matter, $\chi_{\pm}=(\chi^0_1\pm\chi^0_2)/\sqrt{2}$, which will annihilate, among other channels, $\chi_+\chi_+,\chi_+\chi_-,\chi_-\chi_-\rightarrow e^+e^- V$, with $V$ a vector. Since we are interested in the general features from annihilations of $SU(2)$ doublet dark matter particles and not in constructing a fully realistic model, we will assume in what follows that only one of these, $\chi_+$ or $\chi_-$, is present in our Universe today. This can be achieved by postulating a mass splitting between them, so that one of them decays into the other at very early times, for instance by introducing the dimension five operators
\begin{align}
  \begin{split}
    \delta{\cal L}^{\rm fermion}_{\rm mass}&= \frac{1}{\Lambda}\Big[ c_1 (\bar\chi_1 i\sigma_2\Phi^*)(\Phi^\dag i\sigma_2\chi_1^c) +  c_2 (\bar\chi_2 \Phi)(\Phi^T\chi_2^c) + c_3 (\bar\chi_2 \Phi)(\Phi^\dag i\sigma_2\chi_1^c)\Big] + {\rm h.c.}\;,
  \end{split}
\end{align}
with $\Lambda$ a mass scale larger than the Higgs vacuum expectation value and $c_i$ coefficients of order one. Let us denote the dark matter mass eigenstate as $\chi$, the heavier neutral state as $\chi'$ and the charged component by $\chi^\pm$. The mass splittings induced by the dimension five operator are given by,
\begin{eqnarray}
 \delta m_{\pm} & = & m_{\chi^\pm} - m_\chi = \frac{v_{EW}^2}{2\Lambda}(c_3+|c_1-c_2|) \;, \nonumber \\
 \delta m_{0} & = & m_{\chi'} - m_\chi = \frac{v_{EW}^2}{\Lambda}|c_1-c_2| \;,
\end{eqnarray} 
 up to corrections of order $\mathcal{O}(v_{EW}^4/(\Lambda^2 m_\chi))$.
Up to the same order, the mass eigenstates are related to the two doublet fields by $\chi_1=P_L((\chi'+\epsilon\chi)/\sqrt{2},\chi^-)$ and $\chi_2=P_L((\chi^-)^c,(\chi'-\epsilon\chi)/\sqrt{2})$, where $\epsilon={\rm sgn}(c_1-c_2)$, $\chi$ and $\chi'$ are Majorana fields and $\chi^-$ is a Dirac field. We will assume in the following that the radiative corrections to the mass splittings can be neglected compared to the ones induced by the dimension five operators.

By decomposing the gauge interactions of $\chi_1$ and $\chi_2$ into mass eigenstates one obtains
\begin{eqnarray}
 {\cal L}^{\rm gauge}_{\rm int} & = & -\frac{e}{2s_W}\left[\bar\chi \gamma^\mu W_\mu^+ \chi^- + \bar\chi'\gamma_5\gamma^\mu W_\mu^+ \chi^-\right] -\frac{e}{2s_Wc_W}\bar\chi\gamma_5\gamma^\mu Z_\mu\chi' \nonumber\\
& & {} + e \bar\chi^- \gamma^\mu A_\mu \chi^- + e \cot(2\theta_W) \bar\chi^- \gamma^\mu Z_\mu \chi^- \;.
\end{eqnarray}
Note that this coincides with the interactions of the neutralino in the higgsino limit within the MSSM. These interactions give rise to the dark matter annihilation channels into a pair of weak bosons with cross-sections given by
\begin{eqnarray}
\sigma v_{\chi\chi\to WW} & = & \frac{g^4}{32\pi}\frac{m_\chi^2-M_W^2}{(m_\chi^2+m_{\chi^\pm}^2-M_W^2)^2}\sqrt{1-M_W^2/m_\chi^2} \;, \nonumber\\
\sigma v_{\chi\chi\to ZZ} & = & \frac{g^4}{64\pi c_W^4}\frac{m_\chi^2-M_Z^2}{(m_\chi^2+m_{\chi'}^2-M_Z^2)^2}\sqrt{1-M_Z^2/m_\chi^2} \;.
\end{eqnarray}
As is well-known, these cross-sections can be altered substantially for dark matter masses in the TeV range, and if the mass splittings are of order GeV or below, by the multiple exchange of weak bosons among the fermions in the initial state, analogous to Sommerfeld enhancement in electrodynamics \cite{Hisano:2003ec}. We will assume here that the mass splittings are large enough, such that the effect of Sommerfeld enhancement is in a perturbative regime, and comment on its impact below. As we will see this requires $\Lambda\lesssim 10$TeV.

In the present work, we are motivated by the observation that internal bremsstrahlung can lift the helicity suppression of fermionic final states. In fact, as we will discuss below, the annihilation $\chi\chi\to e^+e^-V$ can be  under certain conditions as important as the gauge processes $\chi\chi\to WW, ZZ$. Let us start by discussing the fermionic interactions analogous to the $SU(2)$ singlet case. In particular, the dark matter can couple to the right-handed electron singlet and a scalar field $\eta\equiv(1,2,-\frac{1}{2})$, yielding the two following interaction terms in the Lagrangian:
\begin{align}
  \begin{split}
    {\cal L}^{\rm fermion}_{\rm int}&= f  (\bar\chi_1  i\sigma_2 \eta^*)e_R+
    {\rm h.c.}\;,  \\
    {\cal L}^{\rm scalar}_{\rm int}&= -\lambda_3(\Phi^\dagger \Phi)
    (\eta^\dagger \eta)
    -\lambda_4(\Phi^\dagger \eta)(\eta^\dagger \Phi)\;.
  \end{split}
\end{align}
Alternatively, the dark matter particle can couple to the left-handed electron doublet and a scalar field $\eta\equiv(1,1,1)$,
\begin{align}
  \begin{split}
    {\cal L}^{\rm fermion}_{\rm int}&= 
    f (\bar \chi_1 i\sigma_2 L_e^c) \eta+{\rm h.c.}\;,  \\
    {\cal L}^{\rm scalar}_{\rm int}&= -\lambda_3(\Phi^\dagger \Phi)
    (\eta^\dagger \eta)\;.
  \end{split}
\end{align}   
In a supersymmetric context, the scalars can be identified with $\tilde L_e$ and $\tilde e_R$, respectively.
Note that one could in principle consider additional scalar particles that lead to analogous
couplings involving $\chi_2$. We do not consider this possibility in the following. 

The relic abundance produced by the thermal freeze-out is determined by the $2\to 2$ cross-sections arising
from gauge and Yukawa interactions. If the latter are subdominant, an abundance in accordance with the
WMAP value can be achieved for dark matter masses of order TeV \cite{Cirelli:2005uq}. By adding the
Yukawa interactions, and adjusting the coupling $f$, it is in principle possible to obtain the WMAP value also for
even higher dark matter masses. However, in the following we will not restrict the range of the dark  matter mass or 
the coupling $f$ in order to determine the constraints arising from the measurements of the antiproton flux in a way that is independent of the production mechanism.

We will first discuss the relative size of the electromagnetic and electroweak bremsstrahlung in analogy to the case of singlet dark matter, and then the relative importance of fermionic and diboson final states. Throughout, we will assume that
$m_\eta-m_\chi\gg\delta m_\pm,\delta m_0$, and use the notation $m_{\rm DM}\equiv m_\chi$ in analogy to the singlet case.

\subsubsection*{Coupling to right-handed electrons}

\begin{figure}
\hspace*{-1.5cm}
\begin{tabular}{ll}
 \includegraphics[width=0.55\textwidth]{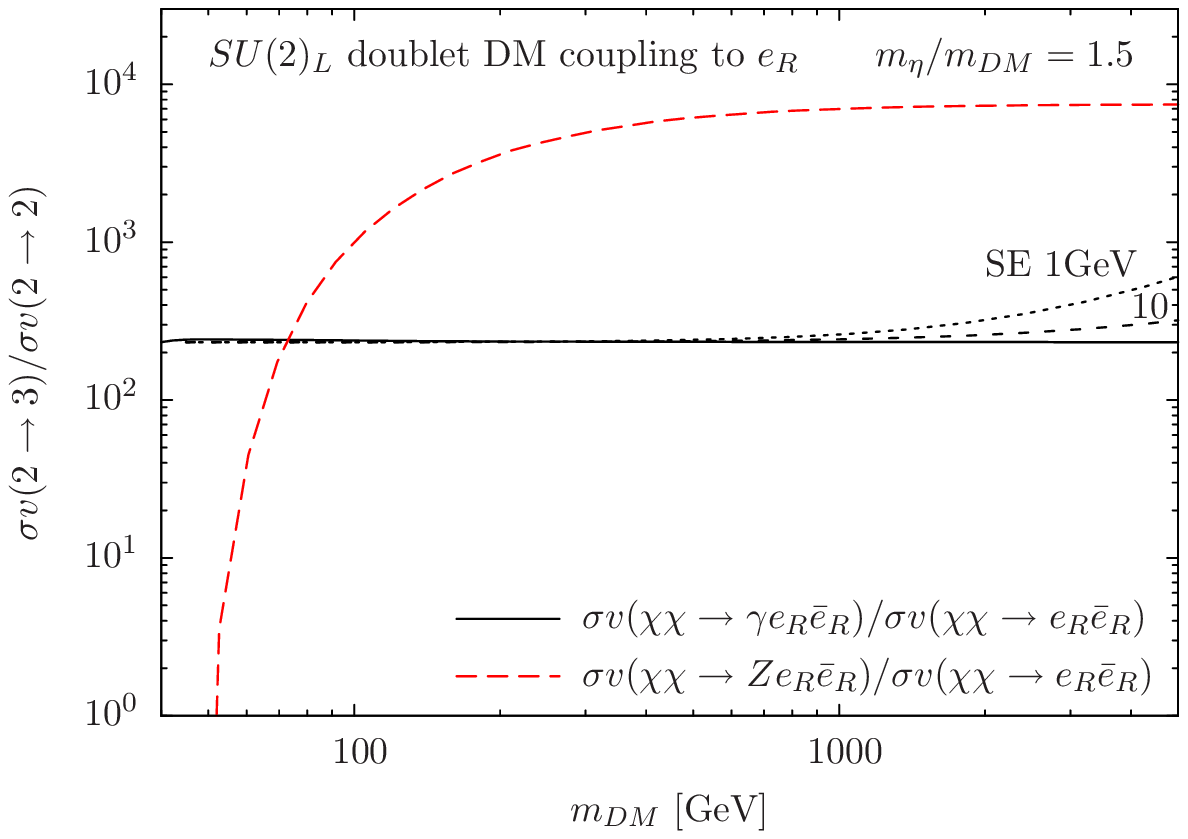}
 &\hspace*{-0.8cm} \includegraphics[width=0.55\textwidth]{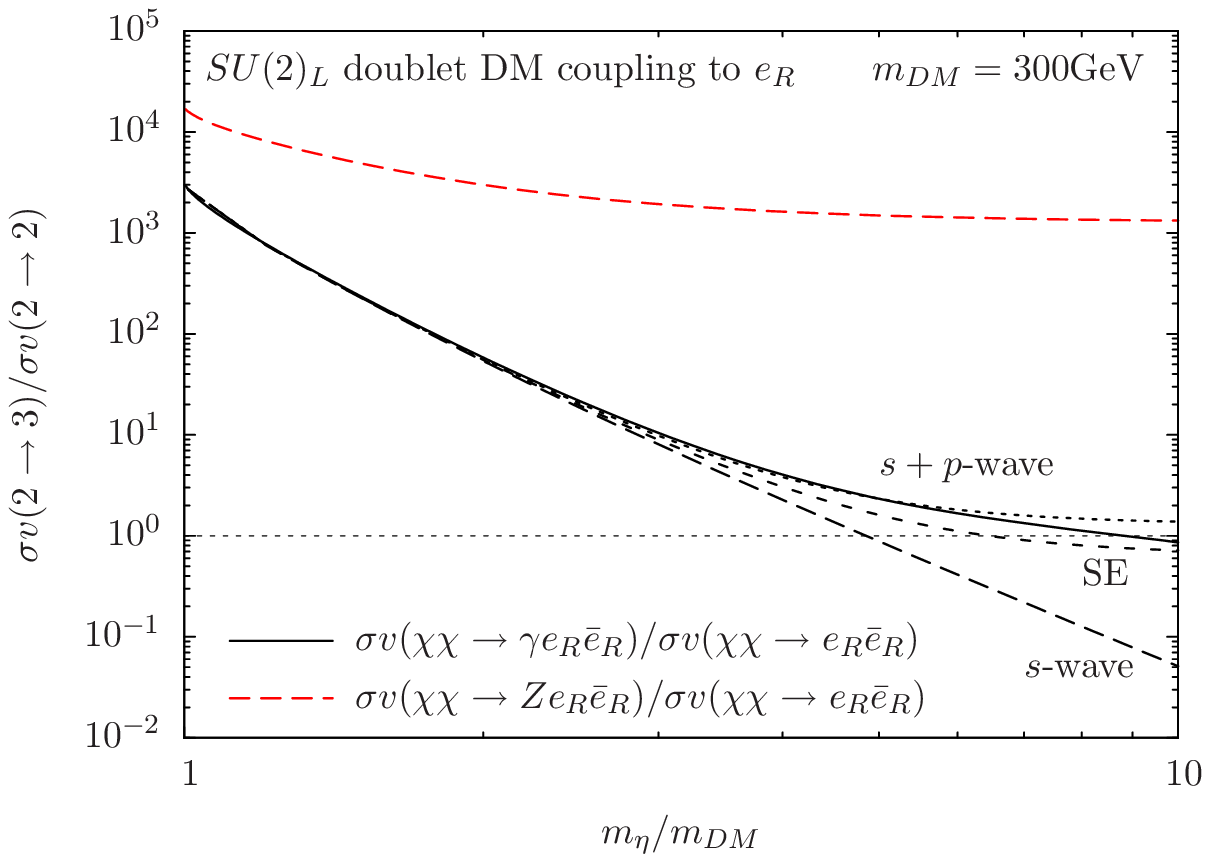} \\
 \includegraphics[width=0.55\textwidth]{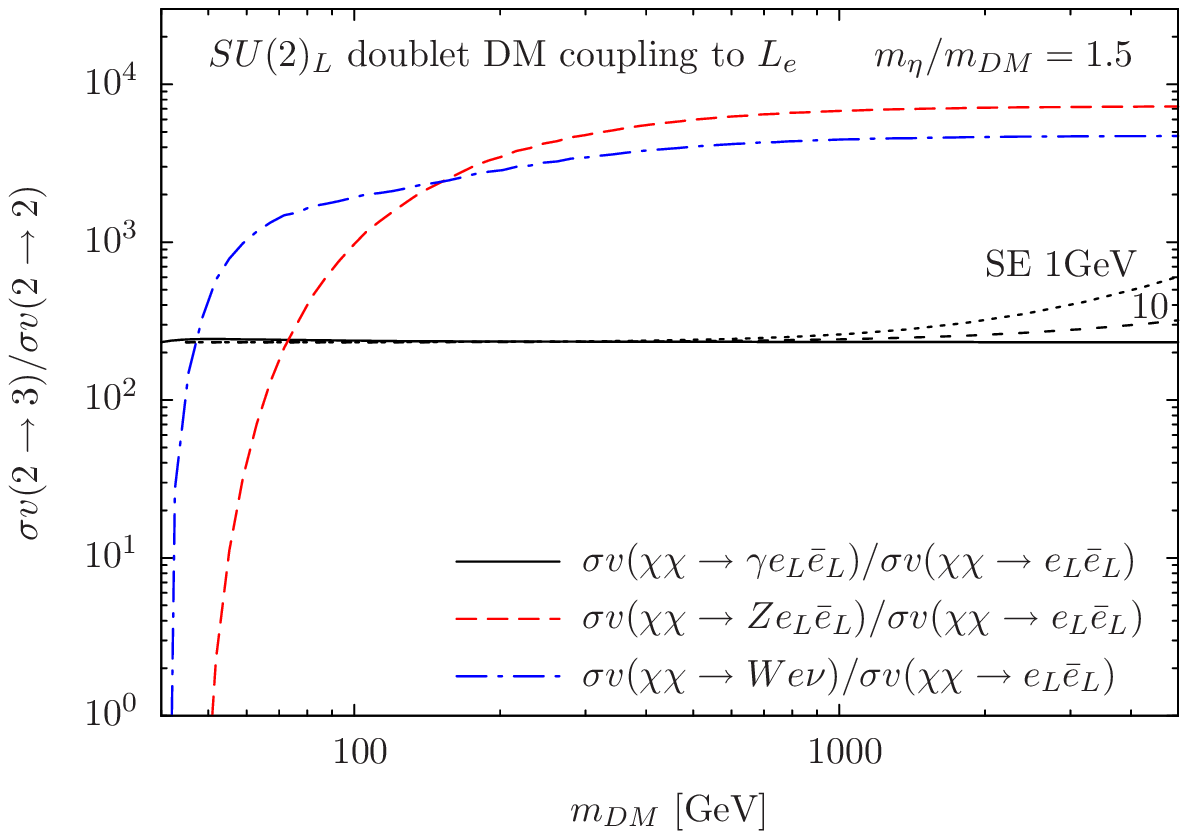}
 &\hspace*{-0.8cm} \includegraphics[width=0.55\textwidth]{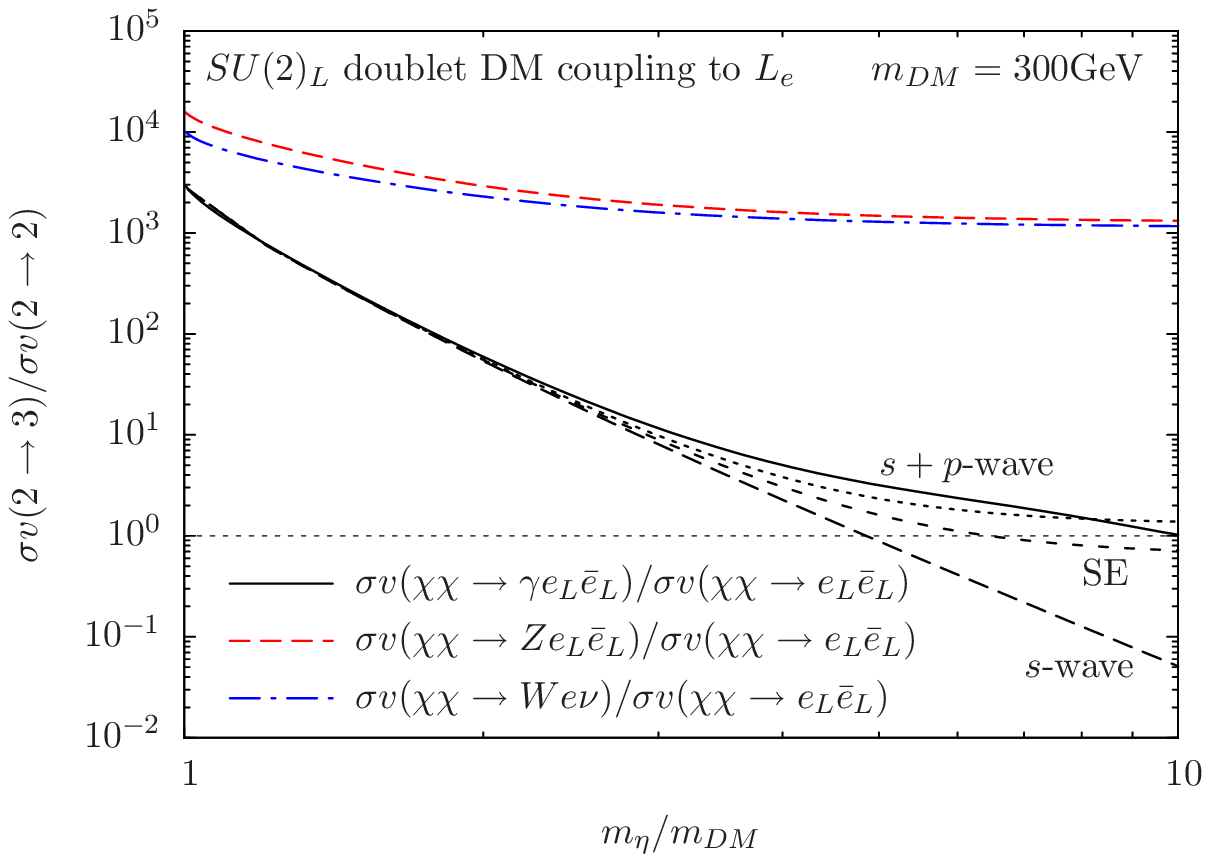} \\
\end{tabular}
 \caption{\label{fig:crossSection1_Doublet_EL} Ratio of three-body and two-body annihilation cross-sections, for the case of
$SU(2)_L$ doublet dark matter coupling to the right-handed electron, and to the left-handed electron doublet,
respectively. The left column shows the dependence on the dark matter mass for fixed ratio $m_{\eta^\pm}/m_{\rm DM}=1.5$,
while the right column shows the dependence on the mass of the mediating scalar particle $\eta$ for $m_{\rm DM}=300$GeV.
Also shown is the correction from Sommerfeld enhancement (SE) for mass splittings $\delta m_\pm=1,10$GeV between
the charged and the lightest neutral components of the dark matter doublets.
For the relative dark matter velocity we use $v=10^{-3}c$.}
\end{figure}

If the mediating scalar $\eta$ has quantum numbers $\eta=(1,2,-\frac{1}{2})$,
it leads to annihilations of dark matter into right-handed electrons.
For $m_{\rm DM}\gg \frac{M_Z}{2}$ and $m_{\eta} \gg m_{\rm DM}$, and taking only annihilations mediated by
the scalar $\eta$ into account, the branching ratio of electromagnetic to electroweak bremsstrahlung is
given approximately by
\begin{equation}
\begin{array}{ccccc}
  \sigma v(\chi\chi\to Z e\bar e) & : & \sigma v(\chi\chi\to \gamma e\bar e) & \simeq & \frac{50\mu(\mu-2s_W^2)+15(1+2s_W^2)^2-3}{60 s_W^2c_W^2}  
\end{array}
\end{equation}
where $\mu=m_{\eta^\pm}^2/m_{\rm DM}^2$, and $s_W$ and $c_W$ are the sine and cosine of the weak mixing angle. More general analytical expressions are given
in the Appendix. The neutral component $\eta^0$ plays no role because it could only lead to
the production of right-handed neutrinos, which are absent in the SM. Note that the branching ratio increases

with the mass of the mediating particle to the fourth power, $\mu^2\propto m_{\eta^\pm}^4$. The reason for this behaviour is that the dark matter particle, being an $SU(2)_L$ doublet, couples also to the $Z$ boson. Therefore, annihilation to
$Z e\bar e$ can occur also via initial state radiation. The latter leads to a non-zero contribution to the s-wave
cross-section already at the $1/\mu^2$-level, while electromagnetic bremsstrahlung occurs at order $1/\mu^4$.
A similar result has been obtained within an effective operator approach for Wino-like
dark matter in \cite{Ciafaloni:2011gv}.
Note that, for very large values of $\mu$, eventually the p-wave contribution to $\gamma e\bar e$ will dominate over the s-wave contribution to $\gamma e\bar e$. However, the former also scales like $1/\mu^2$. This means the ratio of electroweak
and electromagnetic cross sections saturates for very large values of $\mu$,
which can be estimated roughly as $\mu\sim (5 v^2 \ln^2(m_{\rm DM}/m_e))^{-1/2}\sim \mathcal{O}(30)$ for $v=10^{-3}c$
and $m_{\rm DM}\sim 10^2$GeV. However, since a strong gamma signal with spectrum peaked at high energies is produced
only for $\mu\sim \mathcal{O}(1)$, we will not discuss this case in further detail.
The full dependence of the branching ratios on the dark matter mass and the mass of the mediating scalar particle is shown in the upper part of Fig.~\ref{fig:crossSection1_Doublet_EL}. In the right part, it is also shown that the p-wave
contribution to the annihilation into $\gamma e\bar e$ becomes important when $\mu$ is large.

Since the electroweak bremsstrahlung is strongly enhanced compared to electromagnetic bremsstrahlung even for moderate values $\mu\gtrsim 1.5$, it is important to investigate whether higher-order contributions can lift the $1/\mu^4$ suppression of electromagnetic bremsstrahlung. It turns out that this is indeed the case, when considering the corrections arising from
Sommerfeld enhancement. Here, we will estimate the leading effect when the enhancement is perturbatively small, following
Ref.~\cite{Iengo:2009ni,Hryczuk:2011vi}. In general the s-wave amplitude $\mathcal{A}_{\chi\chi\to SM}$ for annihilation into some SM final
state can be written as
\begin{equation}
  \mathcal{A}_{\chi\chi\to SM} = s_0 \mathcal{A}^0_{\chi\chi\to SM} + s_0' \mathcal{A}^0_{\chi'\chi'\to SM}
+ s_\pm \mathcal{A}^0_{\chi^+\chi^-\to SM}
\end{equation}
where the amplitudes $\mathcal{A}^0$ denote the tree-level amplitudes for annihilations of the various components of the doublet, and $s_i\equiv\partial_x\varphi_i(x)|_{x\to 0}$ are enhancement factors. They  are related to the wave-functions
$\varphi_i(x)$ for radially symmetric (s-wave) two-fermion initial states, where $x=r\cdot m_{\rm DM} v/c$ is a dimensionless
variable related to the spatial separation $r$ of the fermions. The wave-functions are solutions of a set of
coupled Schr\"odinger equations in the presence of a  Yukawa potential $\propto e^{-m_V r}/r$ 
that is generated by the exchange of vector bosons $V=W,Z,\gamma$ among the fermion pair~\cite{Iengo:2009ni,Hisano:2002fk}.
For the dark matter masses we are interested in we can safely apply the low-velocity limit
$v/c \sim 10^{-3} \ll M_W/m_{\rm DM}$ and assume that $\mathcal{E}=m_{\rm DM}v^2\ll 2\delta m_{\pm},2\delta m_0$. The latter condition implies that the charged and heavier neutral components of the doublet cannot be produced on-shell, such that their wave-functions decay exponentially at large separations. At leading order in the gauge couplings one then finds the approximate solution
\begin{equation}
s_0 \simeq 1,\ s_0' \simeq \frac{\alpha_{em}}{\sqrt{2}s_{2W}^2}\frac{m_{\rm DM}}{M_Z+\sqrt{2m_{\rm DM}\delta m_0}},\
s_\pm \simeq \frac{\alpha_{em}}{2\sqrt{2}s_{W}^2}\frac{m_{\rm DM}}{M_W+\sqrt{2m_{\rm DM}\delta m_\pm}} \;.
\end{equation}
The approximation can be expected to hold if $|s_0'|,|s_\pm|\ll 1$. For the range of dark matter masses we are
interested in, this is safely the case if $\delta m_{\pm},\delta m_0 \gtrsim \mathcal{O}(1)$GeV$\times (m_{\rm DM}/$TeV$)$
(see e.g. \cite{Hisano:2003ec}). Here, we will assume that this inequality holds and thus that the tree-level
cross sections yield a reliable estimate. Nevertheless, for the annihilation involving electromagnetic bremsstrahlung, the annihilation proceeding via an intermediate charged fermion pair, $\chi\chi\to\chi^+\chi^-\to\gamma e\bar e$, can be important. Concretely, we find that for $m_{\eta} \gg m_{\rm DM}$
\begin{equation}
  \frac{\sigma v(\chi\chi\to\chi^+\chi^-\to \gamma e\bar e)}{\sigma v(\chi\chi \to \gamma e\bar e)} \simeq |s_\pm|^2\frac{\sigma v(\chi^+\chi^-\to \gamma e\bar e)|_{S=0}} {\sigma v(\chi\chi \to \gamma e\bar e)}  \simeq |s_\pm|^2 \frac{50\mu(\mu-1)+57}{15}  \;.
\end{equation}
Here the Sommerfeld enhanced cross section is normalized to the leading order cross-section, and the cross section for $\chi^+\chi^-\to \gamma e\bar e$ should be evaluated for an initial state with total spin zero~\cite{Iengo:2009ni} (note that, for the same reason, the channel $\chi\chi\to \chi^+\chi^-\to e\bar e$ is helicity suppressed although $\chi^\pm$ is a Dirac particle). Thus, as expected, we find that for the annihilation via a charged intermediate state the cross section scales like $1/\mu^2$ instead of $1/\mu^4$ as for the tree-level contribution. Therefore, if $\mu$ is large enough, it may compensate the suppression factor $|s_\pm|^2$, and yield a significant contribution to the annihilation via electromagnetic bremsstrahlung. The influence of Sommerfeld enhancement is shown also in Fig.~\ref{fig:crossSection1_Doublet_EL} for mass splittings $\delta m_\pm=1,10$GeV. Note that analogous corrections exist also for electroweak bremsstrahlung. However, since their leading order cross sections scale already like $1/\mu^2$, these will be small corrections even when $\mu$ is large.

\begin{figure}
\begin{center}
 \includegraphics[width=0.95\textwidth]{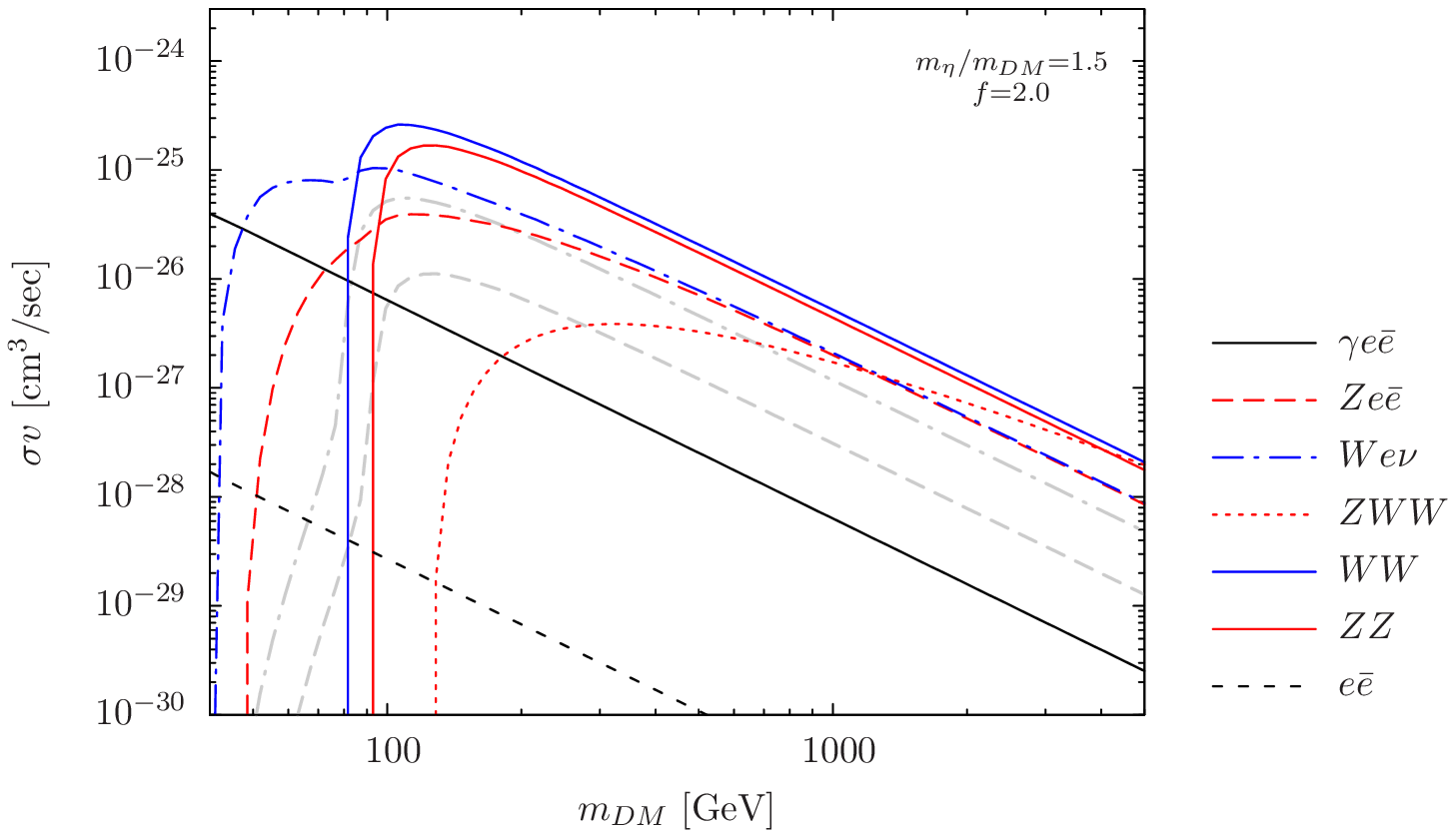} \\
 \includegraphics[width=0.95\textwidth]{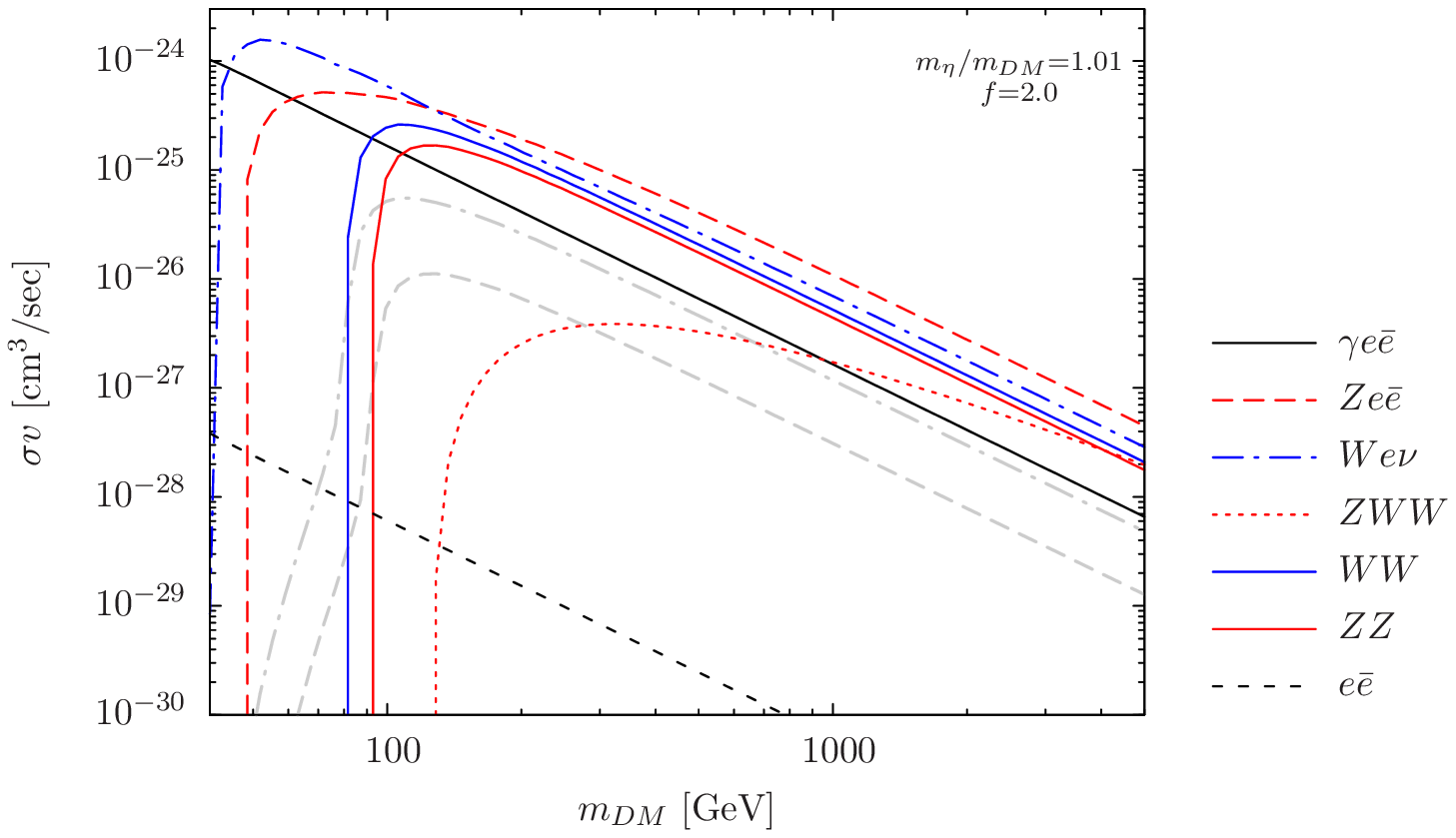}
\end{center}
 \caption{\label{fig:crossSection1_Doublet_EL_Gauge} Cross-sections of various annihilation channels for the case of
$SU(2)_L$ doublet dark matter coupling to the left-handed electron doublet, for $m_{\eta^\pm}/m_{\rm DM}=1.5$ (top)
and $1.01$ (bottom). In both cases, the Yukawa coupling is chosen as $f=2$. The annihilation cross-section into
$We\bar\nu$ and $Ze\bar e$ has two contributions, one from t-channel exchange of the scalar $\eta$ and one from
annihilation into $WW$ or $ZZ$ with a subsequent decay of one of the gauge bosons. The contribution from the latter are shown also as
grey lines. Above the threshold for $WW$/$ZZ$, the latter processes are taken into account already by the $2\to 2$
cross-sections, and are shown here for illustration only. For the relative dark matter velocity we use $v=10^{-3}c$.}
\end{figure}

\subsubsection*{Coupling to left-handed electrons}

If the mediating scalar $\eta$ has quantum numbers $\eta=(1,1,1)$, it leads to annihilations of dark matter into left-handed electrons. For $m_{\rm DM}\gg \frac{M_Z}{2}$ and $m_{\eta} \gg m_{\rm DM}$, we find the following ratios
\begin{equation}
\begin{array}{ccccc}
  \sigma v(\chi\chi\to Z e\bar e) & : & \sigma v(\chi\chi\to \gamma e\bar e) & \simeq & \frac{50\mu(\mu-1+2s_W^2)+60c_W^4-3}{60 s_W^2c_W^2} \\
  \sigma v(\chi\chi\to W e\nu) & : & \sigma v(\chi\chi\to \gamma e\bar e)
  & \simeq & \frac{50\mu(\mu-1)+63}{60s_W^2}  
\end{array}
\end{equation}
Annihilation into $Z\nu\bar \nu$ does not occur at tree level, because the quantum numbers of $\eta$ imply
that it couples the dark matter particle only to the charged lepton. Also, since $\eta$ is a $SU(2)_L$ singlet, only transversally polarized $W$ bosons are produced via initial as well as final state radiation provided that $\delta m_\pm,\delta m_0\ll M_W$. The reason for the
$\mu^2$-dependence of the ratios of cross sections is due to initial state radiation, as discussed above.
The full dependence of the branching ratios on the dark matter mass and the mass of the mediating scalar particle is shown in the lower part of Fig.~\ref{fig:crossSection1_Doublet_EL}.

When including Sommerfeld corrections, the annihilation channels $Z\nu\bar\nu$ and $\gamma\nu\bar\nu$ appear. In particular,
the latter yields a contribution that scales like $1/\mu^2$ instead of $1/\mu^4$ as for $\gamma e\bar e$. Its cross section
is given by
\begin{equation}
  \frac{\sigma v(\chi\chi\to\chi^+\chi^-\to \gamma \nu\bar \nu)}{\sigma v(\chi\chi \to \gamma e\bar e)} \simeq |s_\pm|^2\frac{\sigma v(\chi^+\chi^-\to \gamma \nu\bar \nu)|_{S=0}} {\sigma v(\chi\chi \to \gamma e\bar e)}  \simeq |s_\pm|^2 \frac{50\mu^2+12}{15}  \;.
\end{equation}
The Sommerfeld corrections shown in the lower panel of Fig.~\ref{fig:crossSection1_Doublet_EL} refer to
the sum of the electromagnetic annihilation cross sections $\sigma v(\chi\chi \to \gamma e\bar e)+\sigma v(\chi\chi\to\chi^+\chi^-\to \gamma \nu\bar \nu)$.

\medskip

The relative size of the cross sections $\chi\chi\to WW,ZZ$ and $\chi\chi\to Ve\bar e,V\nu\bar\nu,We\bar\nu$ depend on the
ratio of the Yukawa coupling $f$ to the gauge coupling $g$ as well as the ratio of the mass of the mediating
particle $\eta$ and the dark matter mass. In general, due to the lifting of helicity suppression, the $2\to 3$
processes are much less suppressed than the corresponding $2\to 2$ annihilations into fermions. However, the
$2\to 2$ annihilations mediated by gauge interactions are also not helicity suppressed and therefore generically
dominate over the $2\to 3$ channels. Nevertheless, there can be some cases when the latter are important.
For example, this can be the case if the dark matter mass is in the range $M_W\lesssim m_{\rm DM}\lesssim 2M_W$,
such that the diboson states are kinematically disfavored. Another possibility is a rather large value for
the coupling $f$. In Fig.~\ref{fig:crossSection1_Doublet_EL_Gauge} the various cross sections are shown
for $f=2$. In this case, the cross sections of the $2\to 3$ channels arising from electroweak bremsstrahlung
are the dominant annihilation channels provided that $m_\eta/m_{\rm DM}\lesssim 1.01$. For even larger $f$, the latter
restriction can be relaxed. We note that couplings of that size are required for dark matter masses in the
multi-TeV range, when imposing the relic density constraint from thermal freeze out. In order to determine
constraints from the antiproton flux, we will consider both the case that the annihilation channels
$\chi\chi\to WW,ZZ$ or $\chi\chi\to Ve\bar e, We\bar\nu$ dominate.

\section{Dark matter coupling to quarks}\label{sec:quarks}

The analysis of the annihilation of dark matter particles into quarks is completely parallel to the annihilation into leptons discussed in the previous section, the main difference being the inclusion of a color quantum number in the final fermion states and in the intermediate scalar state. As a result, in addition to the electromagnetic and electroweak bremsstrahlung processes, a new annihilation channel arises where a gluon can be radiated off the final quark states or off the internal colored scalar state. Being both the gluon and the photon massless gauge bosons, the resulting spectra will be identical \cite{Flores:1989ru}. However, the cross section for the gluon internal bremsstrahlung will be enhanced compared to the electromagnetic internal bremsstrahlung by the larger coupling constant and by the color factor, therefore we expect a larger impact of the $2\rightarrow 3$ processes in scenarios where the dark matter couples to quarks compared to scenarios where the dark matter couples to leptons, and in particular a larger impact of the present measurements of the antiproton-to-proton fraction on the constraints on the couplings of the model.

Scenarios where the dark matter particle couples to quarks and to a colored scalar particle are strongly constrained by experiments searching for exotic colored particles. The LEP constraint on the invisible width of the $Z$ boson, $\Delta \Gamma_{\rm inv}<2.0$ MeV~\cite{:2005ema}, allows to set the absolute lower limits $m_{\eta}>44$ GeV if $\eta\equiv (3,2,\frac{1}{6})$ or $(3,2,\frac{2}{3})$ and $m_{\eta}>33$ GeV if  $\eta\equiv (3,2,-\frac{1}{3})$, corresponding to the searches for left-handed quark doublets, right-handed up quarks and right-handed down quarks, respectively~\cite{Nakamura:2010zzi}.

Recently, the ATLAS collaboration has reported in \cite{Aad:2011ib} the results for the search of squark and gluinos using final states with jets and missing transverse momentum using 1.04 fb$^{-1}$ of data taken in proton-proton collisions with $\sqrt{s}=7$ TeV at the Large Hadron Collider. In this analysis it is considered a simplified supersymmetric model with R-parity conserved containing only squarks of the first two generations, $\tilde q$, a gluino octet, $\tilde g$, and the lightest neutralino, $\tilde  \chi_1^0$, while all other supersymmetric particles are assumed to be very heavy. In this simplified scenario, the supersymmetric particles are produced in pairs and the final states produce more than 2, 3 or 4 jets plus missing energy. Concretely, the production of two squarks $\tilde q\tilde q$ is followed by the decay $\tilde q\rightarrow q \tilde \chi_1^0$, thus producing at least two jets plus missing energy. In contrast the production of two gluinos $\tilde g \tilde g$ is followed by the decay $\tilde g\rightarrow q\bar q \chi_1^0$, which yields in the final state at least four jets plus missing energy. Lastly, the associated production of one squark and one gluino $\tilde q\tilde g$ yields at least three jets plus missing energy. The non observation of an excess in any of these channels above the Standard Model background can be translated into constraints on the $(m_{\tilde g},m_{\tilde q})$ plane, assuming $m_{\chi_1^0}=0$.

Our toy model for dark matter corresponds to the simplified SUSY model considered by the ATLAS collaboration in the limit where the gluino is also very heavy and is not kinematically accessible to the LHC with  $\sqrt{s}=7$ TeV. Hence, we conclude that the non-observation of an excess over the Standard Model background of dijet events with missing energy in the present LHC data translates into a lower bound on the colored scalar state of our toy model $m_{\eta}>875$ GeV at 95\% c.l. It is important to note that this lower bound assumes that the mass splitting between $\chi$ and $\eta$ is large enough to produce jets passing the requirements for the transverse momenta ($p_T>130$ GeV for the leading jet and $p_T>40$ GeV for the second jet). Then, this stringent lower bound on the colored scalar mass can be avoided if $\chi$ and $\eta$ present a degenerate mass spectrum,  as discussed in~\cite{Kawagoe:2006sm}, concretely when $ m_{\eta}-m_{\rm DM}< 130$ GeV, so that the dijet event does not pass all the cuts required by the ATLAS analysis. 

Searches for colored scalar states were also undertaken at the Tevatron and at LEP. The searches at the Tevatron by the CDF~\cite{Aaltonen:2008rv} and D0~\cite{:2007ww} collaborations employ similar cuts as the ATLAS analysis described above, and have by now been superseded. On the other hand, the searches at LEP, despite limited by the smaller center of mass energy and by the smaller luminosity, employed a smaller cut for the jet transverse momentum and are relevant for our analysis. Concretely, the L3 collaboration has presented limits on the squark masses searching for an excess in dijet events with missing energy in $e^+ e^-$ collisions at center of mass energies between 192 GeV and 209 GeV with an integrated luminosity of 450.5 pb$^{-1}$~\cite{Achard:2003ge}. The non-observation of an excess with respect to the expected Standard Model background translates in our toy model into the lower bound $m_{\eta}\geq 97$ GeV for $m_{\eta}-m_{\rm DM}> 10$ GeV. 

As a summary, we conclude that the toy model with a Majorana dark matter particle which couples to the quarks and a colored scalar, $\eta$, via a Yukawa coupling is in agreement with the present searches of new physics if
\begin{itemize}
\item  $m_{\eta}>875~{\rm GeV}$ for any $m_{\rm DM}$.
\item  $97~{\rm GeV}\leq m_{\eta}\leq 875~{\rm GeV}$, 
if $m_{\eta}-m_{\rm DM}<130~{\rm GeV}$.
\item $33-44~{\rm GeV}\leq m_{\eta}\leq 97~{\rm GeV}$, 
if $m_{\eta}-m_{\rm DM}<10~{\rm GeV}$.
\end{itemize}

Following the same scheme as in the case of annihilations into leptons, we will analyze the features of various dark matter scenarios coupling to quarks according to the charge of the dark matter particle under $SU(2)_L$.

\subsection{$SU(2)_L$ singlet dark matter}

This choice of the $SU(2)_L$ charge requires that the gauge quantum numbers of the dark matter particle must be $\chi\equiv (1,1,0)$ in order to render an electrically neutral particle, while the quantum numbers of the intermediate colored scalar $\eta$ depend on whether the dark matter particle couples to the right-handed up quarks, $u_R=(3,1,\frac{2}{3})$, the right-handed down quarks, $d_R=(3,1,-\frac{1}{3})$, or to the quark doublet, $q_L= \binom{u_L}{d_L}\equiv(3,2,\frac{1}{6})$. 

The fermionic interaction term when the dark matter couples to the  right-handed up quark reads:
\begin{align}
  \begin{split}
    {\cal L}^{\rm fermion}_{\rm int}&= -f \bar \chi u_R \eta+{\rm h.c.}\;,  \\
  \end{split}
\end{align}
which requires quantum numbers for the intermediate scalar particle $\eta\equiv(3,1,-\frac{2}{3})$, while the scalar interaction term is given by Eq.(\ref{eq:singlet-eR}). Similarly, a Yukawa interaction between the dark matter and the down quark requires $\eta\equiv(3,1,\frac{1}{3})$. In the MSSM these particles correspond to the right-handed up and down squarks, respectively.

On the other hand, the fermionic interaction Lagrangian of the dark matter particle to the quark doublet reads:
\begin{align}
  \begin{split}
    {\cal L}^{\rm fermion}_{\rm int}&= -f \bar \chi(q_L i\sigma_2 \eta)+{\rm h.c.}=
    -f \bar \chi (u_{L} \eta^u-d_L \eta^d)+{\rm h.c.}\;, 
  \end{split}
\end{align}
with $\eta=\binom{\eta^u}{\eta^d}\equiv (3, 2,-\frac{1}{6})$, which in the MSSM corresponds to the left-handed squark doublet. The scalar interaction Lagrangian is given in Eq.(\ref{eq:singlet-L}).

In this case, the helicity suppression of the $2\rightarrow 2$ annihilation channels $\chi\chi\rightarrow u_R\bar u_R,~d_R \bar d_R,~q_L \bar q_L$, can be lifted by the associated emission of photons, weak gauge bosons or gluons together with the light quarks. Let us analyze the relative strength of these channels for each of the scenarios.

\subsubsection*{Coupling to right-handed up quarks}

In this case the particle mediating the dark matter annihilations carries hypercharge, electric and color charge, therefore the annihilation channels $\gamma u_R\bar u_R$, $Z u_R\bar u_R$, $g  u_R\bar u_R$ are possible. We show in Fig.~\ref{fig:crossSection1_Q}, upper plot, the corresponding cross sections relative to the helicity suppressed cross section for $\chi\chi\rightarrow u_R \bar u_R$. In the left plot we present the ratio of cross sections for dark matter masses between 50 GeV and 5 TeV and different values of the mass splitting $m_\eta-m_{\rm DM}=100$ GeV, 50 GeV and 10 GeV, to study the impact of the constraints from collider searches of exotic colored particles. It is apparent from the plot that the smaller the mass splitting, the larger is the relative cross section of the $2\rightarrow 3$ processes, especially for light dark matter particles. On the other hand, in the limit $m_{\rm DM}\gg \frac{M_Z}{2}$ the ratios are fairly independent of the mass splitting and take the values:
\begin{equation}
\begin{array}{ccccccc}
  \sigma v(\chi\chi\to g u_R\bar u_R) & : & \sigma v(\chi\chi\to \gamma u_R\bar u_R) & = & 3\alpha_s(m_{\rm DM})/\alpha_{em} & \simeq & 38.4\;, \\
  \sigma v(\chi\chi\to Z u_R\bar u_R) & : & \sigma v(\chi\chi\to \gamma u_R\bar u_R) & = & \tan^2(\theta_W) & = & 0.30\;. \\
\label{eq:ratios-uR}
\end{array}
\end{equation}
Exact formulae for the various cross sections can be found in the Appendix. For the numerical values given here and below,
we have evaluated the strong coupling constant at a scale $m_{\rm DM}=300$GeV for illustration.

Moreover, we show in the upper-right plot the dependence of the ratio of cross sections on the mass of the intermediate colored scalar particle for a fixed dark matter mass $m_{\rm DM}=300$ GeV. The maximal value of the ratio of cross sections, which is as large as $4.6\times 10^4$ for $\chi\chi\rightarrow g u_R \bar u_R$, is reached when $m_\eta\simeq m_{\rm DM}$. We note that this is precisely the region of the parameter space which is most difficult to constrain at colliders, as the jet produced in the decay of the colored scalar particle is too soft to be triggered. As in the leptonic case, for $\mu\equiv m_\eta^2/m_{\rm DM}^2\gg 1$, the cross sections of the $2\rightarrow 3$ processes scale as $1/\mu^4$, while that of the $2\rightarrow 2$ process scale as $1/\mu^2$.

\subsubsection*{Coupling to right-handed down quarks}

The results for the annihilations into right-handed down quarks are completely analogous to the results for the annihilations into right-handed up quarks presented above, the only difference being the different hypercharge (and electric charge) of the intermediate scalar. As a consequence, the cross sections for the annihilations $\chi\chi\rightarrow (Z,\gamma) d_R\bar d_R$ are a factor of 1/4 smaller than the corresponding cross sections for $\chi\chi\rightarrow (Z,\gamma) u_R\bar u_R$. In particular, when $m_{\rm DM}\gg \frac{M_Z}{2}$ the cross sections for annihilations into right-handed down quarks satisfy the relations:
\begin{equation}
\begin{array}{ccccccc}
  \sigma v(\chi\chi\to g d_R\bar d_R) & : & \sigma v(\chi\chi\to \gamma d_R\bar d_R) & = & 12\alpha_s(m_{\rm DM})/\alpha_{em} & = & 154\;, \\
  \sigma v(\chi\chi\to Z d_R\bar d_R) & : & \sigma v(\chi\chi\to \gamma d_R\bar d_R) & = & \tan^2(\theta_W) & = & 0.30\;. \\
\label{eq:ratios-dR}
\end{array}
\end{equation}

\subsubsection*{Coupling to left-handed quarks}

\begin{figure}
\hspace*{-1.5cm}
\begin{tabular}{ll}
 \includegraphics[width=0.55\textwidth]{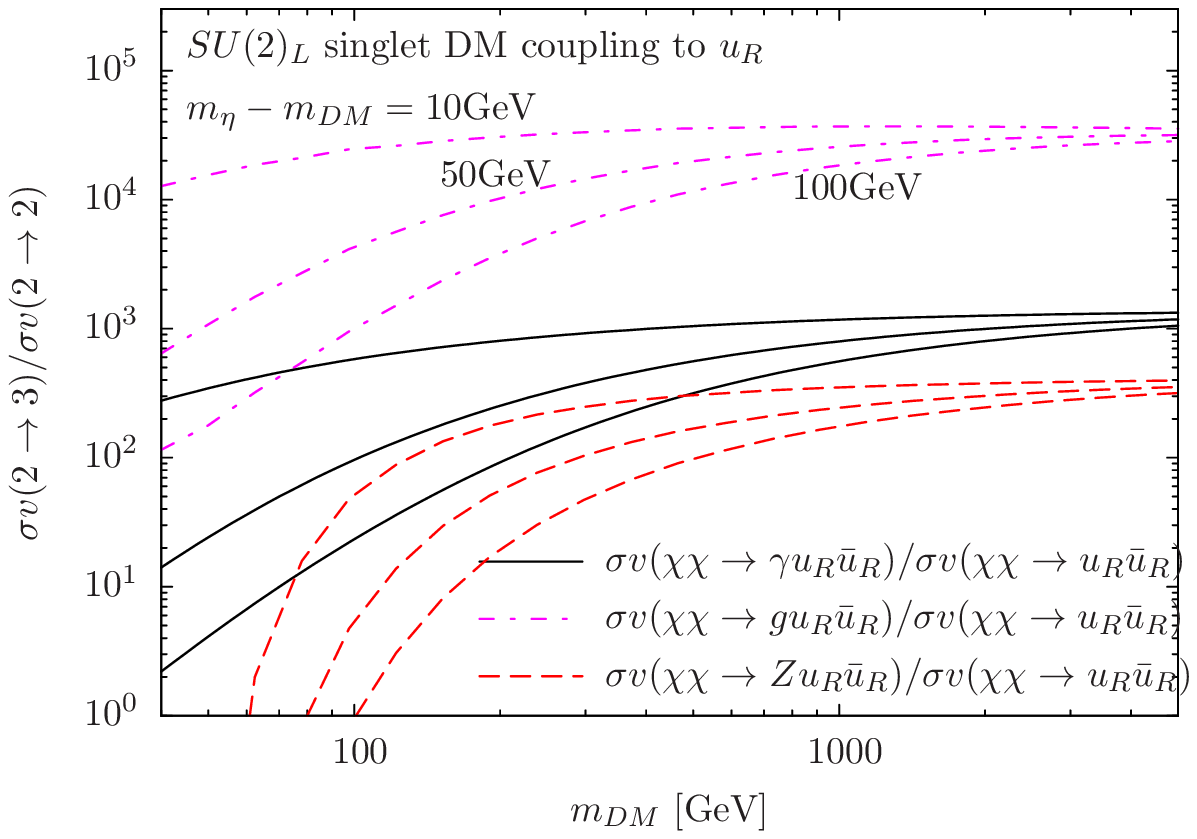}
 &\hspace*{-0.8cm} \includegraphics[width=0.55\textwidth]{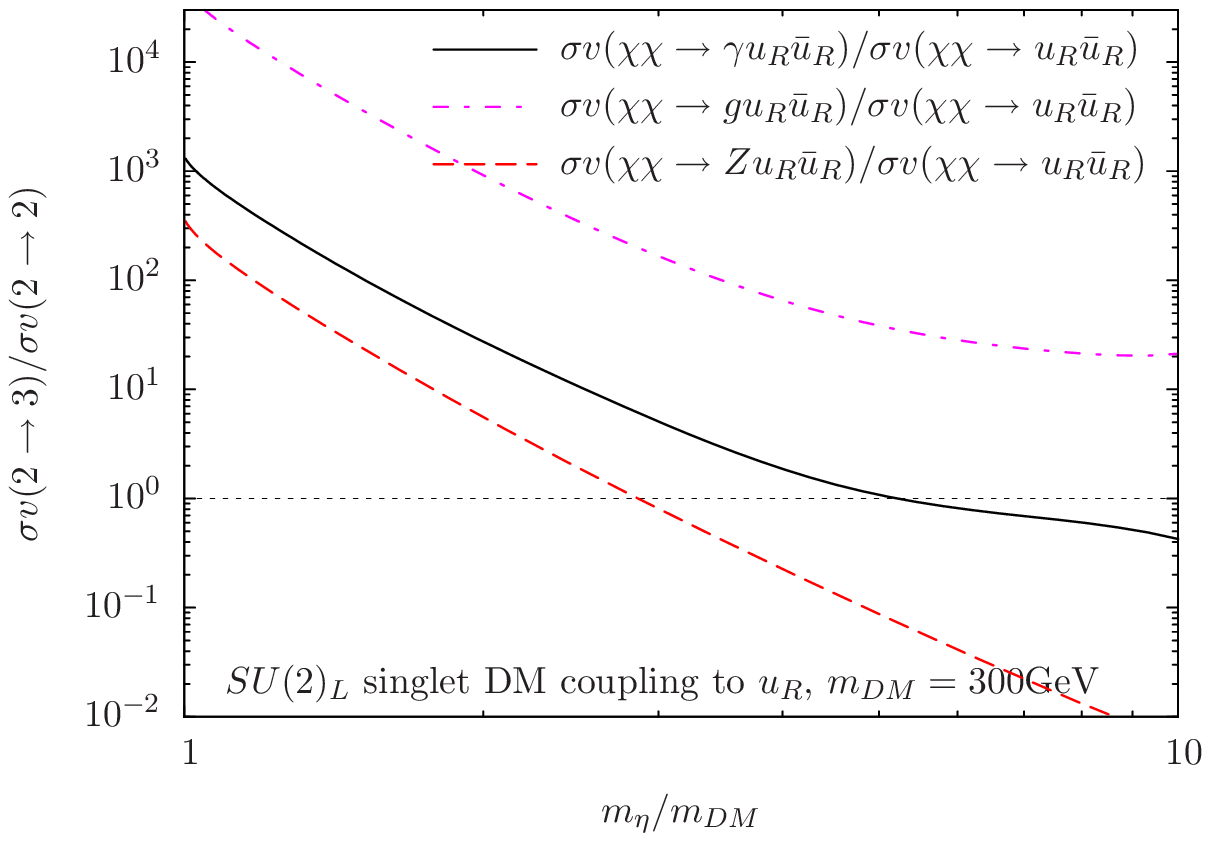} \\
 \includegraphics[width=0.55\textwidth]{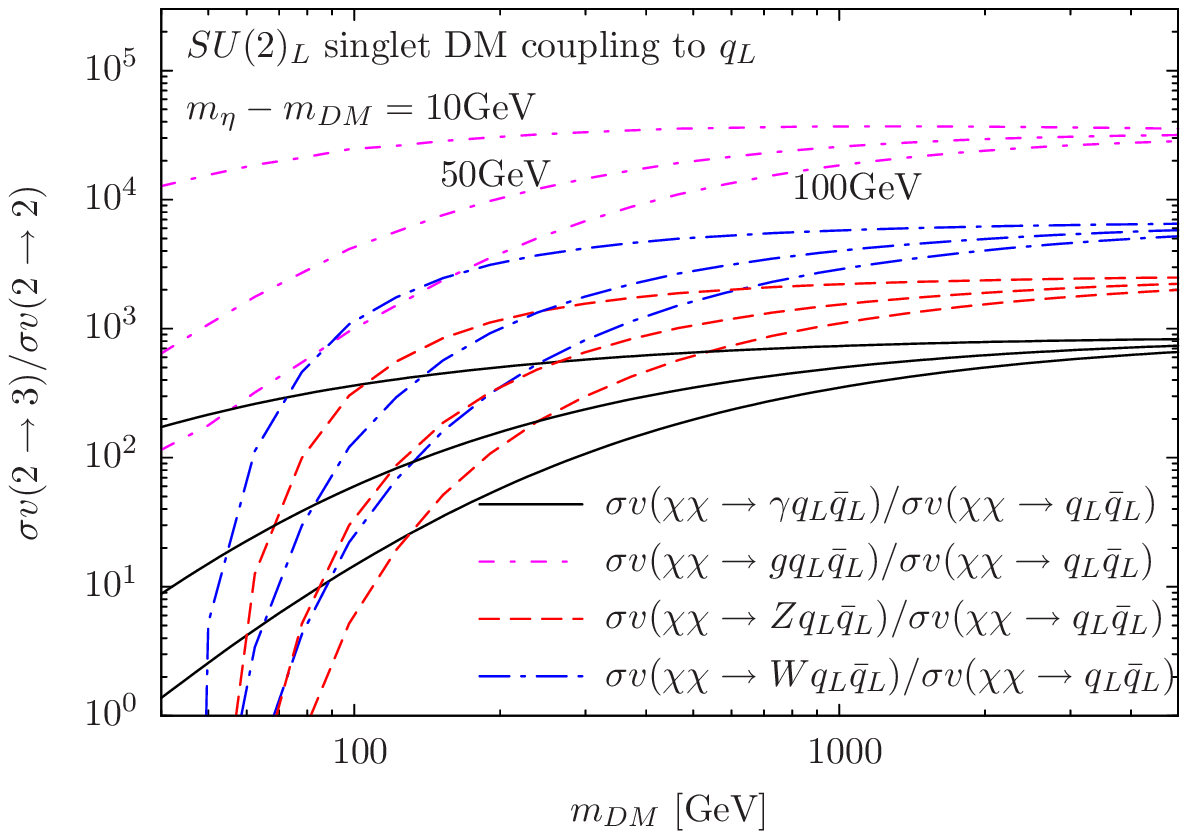}
 &\hspace*{-0.8cm} \includegraphics[width=0.55\textwidth]{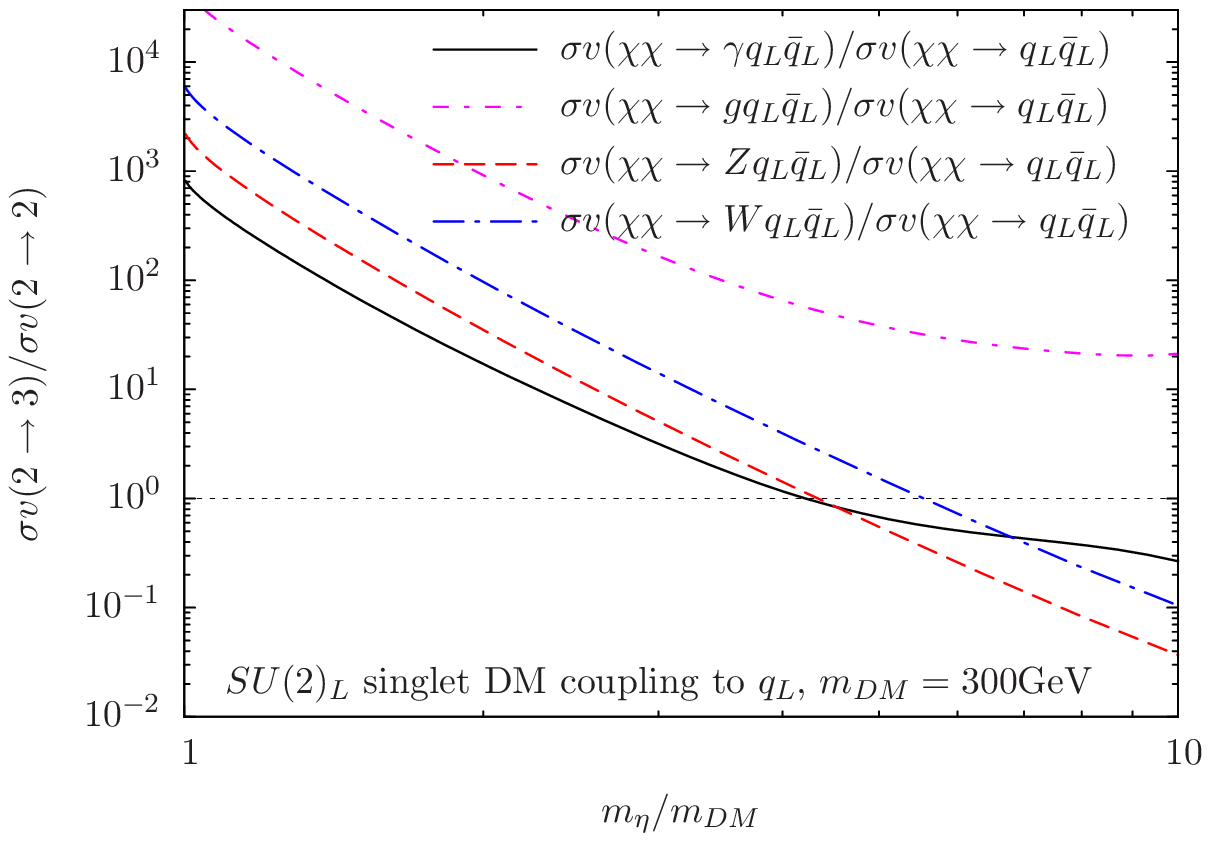} \\
\end{tabular}
 \caption{\label{fig:crossSection1_Q} Ratio of three-body and two-body annihilation cross-sections. The top and bottom rows show the case of $SU(2)_L$ singlet dark matter coupling to the right-handed up-quark, and to the left-handed quark doublet, respectively. The left column shows the dependence on the dark matter mass for fixed mass splitting $m_{\eta}-m_{\rm DM}=10, 50$ and $100$GeV, while the right column shows the dependence on the mass of the mediating scalar particle $\eta$ for $m_{\rm DM}=300$GeV. For the relative dark matter velocity we use $v=10^{-3}c$.}
\end{figure}

When the dark matter particle couples to the left-handed quarks, a new annihilation channel is open, involving $W$ bosons which can be radiated off the internal colored scalar or off the final fermions legs. The ratios $\sigma v(\chi\chi\rightarrow V q_L \bar q_L)/\sigma v(\chi\chi\rightarrow q_L \bar q_L)$ are shown in Fig.~\ref{fig:crossSection1_Q}, lower-left plot, in the limiting case $m_\eta^0=m_{\eta^\pm}$ for dark matter masses between 50 GeV and 5 TeV and different mass splittings between the dark matter mass and the intermediate scalar mass, $m_{\eta}-m_{\rm DM}=$ 10, 50 and 100 GeV. Compared to the coupling into right-handed quarks, it is noticeable the enhancement of the branching ratio into electroweak gauge bosons, arising from the additional channel involving $W$ bosons and both up and down quarks, as well as a stronger coupling of left-handed compared to right-handed quarks to the $Z$-boson. On the other hand, the branching ratio into gluons is identical to the right-handed case, while the branching ratio for $\gamma q_L \bar q_L$ differs by a factor of 8/5 (2/5) with respect to $\gamma u_R\bar u_R$ ($\gamma d_R\bar d_R$) due to the different electric charges of the particles involved and due to the doubling of diagrams. In the limit $m_{\rm DM}\gg \frac{M_Z}{2}$ the cross sections satisfy the relations:
\begin{equation}
\begin{array}{cccclcc}
  \sigma v(\chi\chi\to g q\bar q) & : & \sigma v(\chi\chi\to \gamma q\bar q) & = & 24\alpha_s(m_{\rm DM})/(5\alpha_{em}) & = &  61.4\;, \\
  \sigma v(\chi\chi\to Z q\bar q) & : & \sigma v(\chi\chi\to \gamma q\bar q) & = & \frac{(3-4s_W^2)^2+(3-2s_W^2)^2}{5\sin^2(2\theta_W)}  & = & 3.02 \;,\\
  \sigma v(\chi\chi\to W q\bar q') & : & \sigma v(\chi\chi\to \gamma q\bar q)
  & = & 9/(5s_W^2)  & = & 7.79\;,
\label{eq:ratios-qL}
\end{array}
\end{equation}
with $ \sigma v(\chi\chi\to W q\bar q')= \sigma v(\chi\chi\to W^+ d_L\bar u_L)+
 \sigma v(\chi\chi\to W^- u_L\bar d_L)$.

The dependence of the branching ratios with the mass of the intermediate scalar $\eta$ is shown in  Fig.~\ref{fig:crossSection1_Q}, lower-right plot.

In a more realistic scenario, the two weak isospin components of the intermediate scalar particle will not be degenerate in mass, but will have a mass splitting proportional to the order parameter of the electroweak symmetry breaking: $m^2_{\eta^0}-m^2_{\eta^\pm}=\lambda_4 v^2_{\rm EW}$. As already discussed for dark matter particles coupling to leptons, in this situation the $t$-channel propagators in the annihilation diagrams have poles at different masses and, more importantly, the new channel with annihilations into longitudinally polarized W-bosons opens up, resulting in an enhancement of the branching ratio for $\chi\chi\rightarrow W q_L q_L$. In the limit $m_{\eta^i}\gg m_{\rm DM}\gg \frac{M_Z}{2}$ we find that the cross sections for the $2\rightarrow 3$ annihilation cross sections satisfy:
\begin{eqnarray}
 \frac{ \sigma v(\chi\chi\to g q\bar q) }{ \sigma v(\chi\chi\to \gamma q\bar q) } & = & 12\frac{\alpha_s(m_{\rm DM})}{\alpha_{em}}\frac{\mu_u^{-4}+\mu_d^{-4}}{4\mu_u^{-4}+\mu_d^{-4}} \;, \nonumber\\
 \frac{ \sigma v(\chi\chi\to Z q\bar q)  }{  \sigma v(\chi\chi\to \gamma q\bar q) }& = & \frac{(3-4s_W^2)^2\mu_u^{-4}+(3-2s_W^2)^2\mu_d^{-4}}{\sin^2(2\theta_W)(4\mu_u^{-4}+\mu_d^{-4})}   \;, \\
  \frac{\sigma v(\chi\chi\to W q\bar q')  }{  \sigma v(\chi\chi\to \gamma q\bar q)}
  & = & \frac{9((\mu_u+\mu_d)/2)^{-4}}{s_W^2(4\mu_u^{-4}+\mu_d^{-4})}  \left[ 1 + \frac{5}{8}\frac{m_{\rm DM}^2}{ M_W^2}(\mu_u-\mu_d)^2  \right] \;, \nonumber
\end{eqnarray}
where $\mu_i=m_{\eta^i}^2/m_{\rm DM}^2$. Analytical expressions for the differential cross sections can be found in the Appendix.

\begin{figure}
\hspace*{-1.5cm}
\begin{tabular}{ll}
 \includegraphics[width=0.55\textwidth]{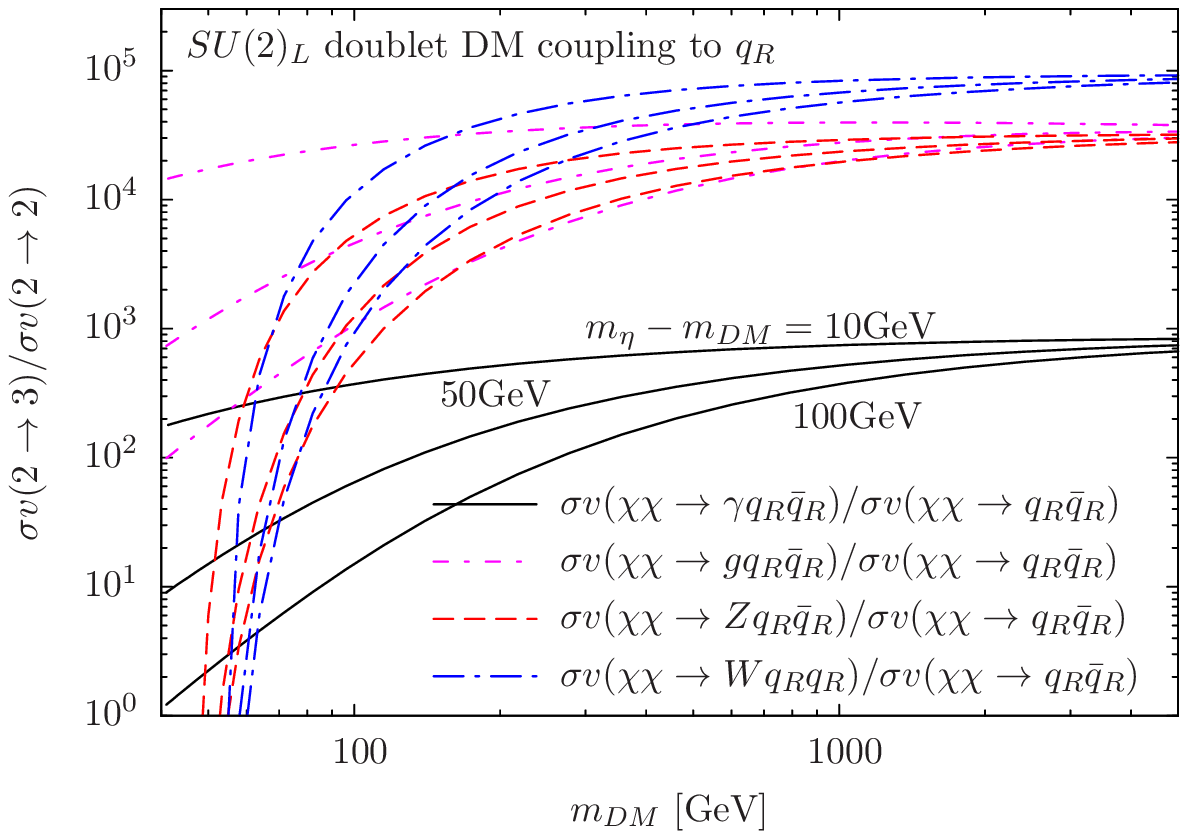}
 &\hspace*{-0.8cm} \includegraphics[width=0.55\textwidth]{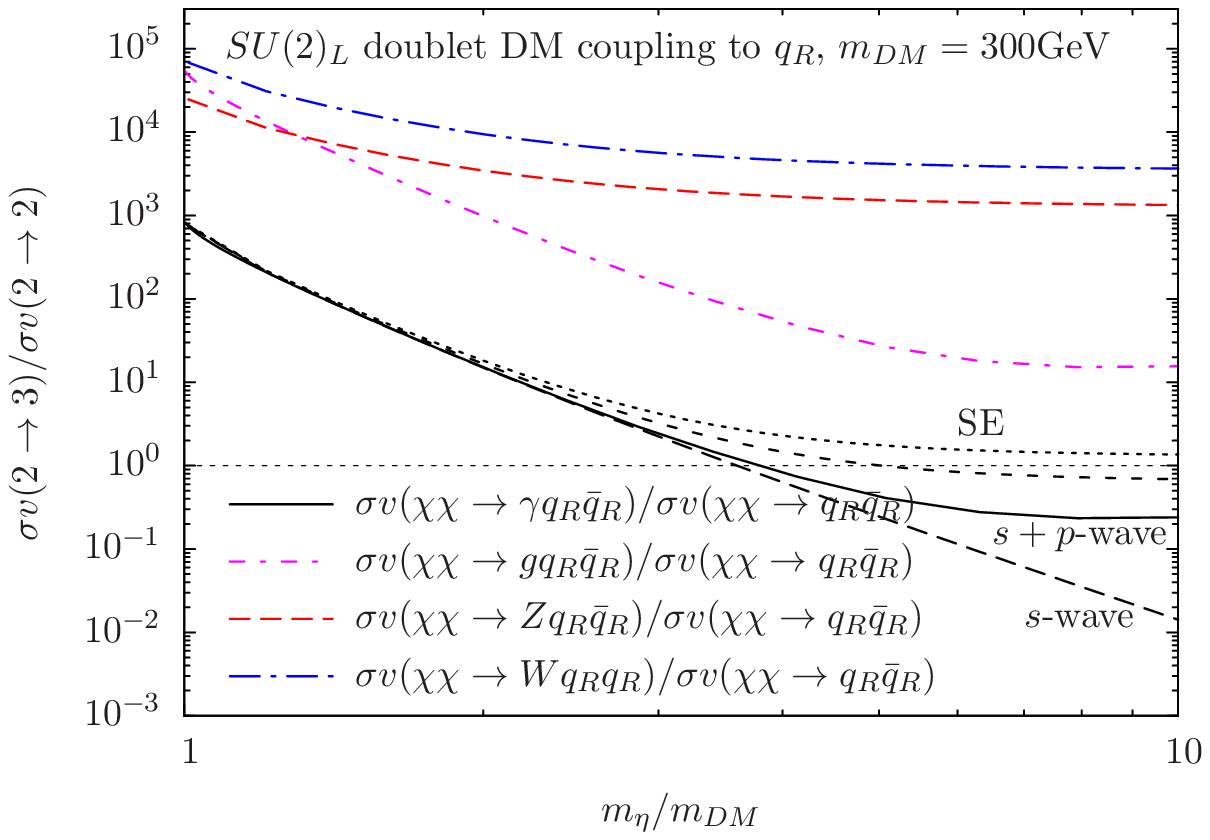} \\
 \includegraphics[width=0.55\textwidth]{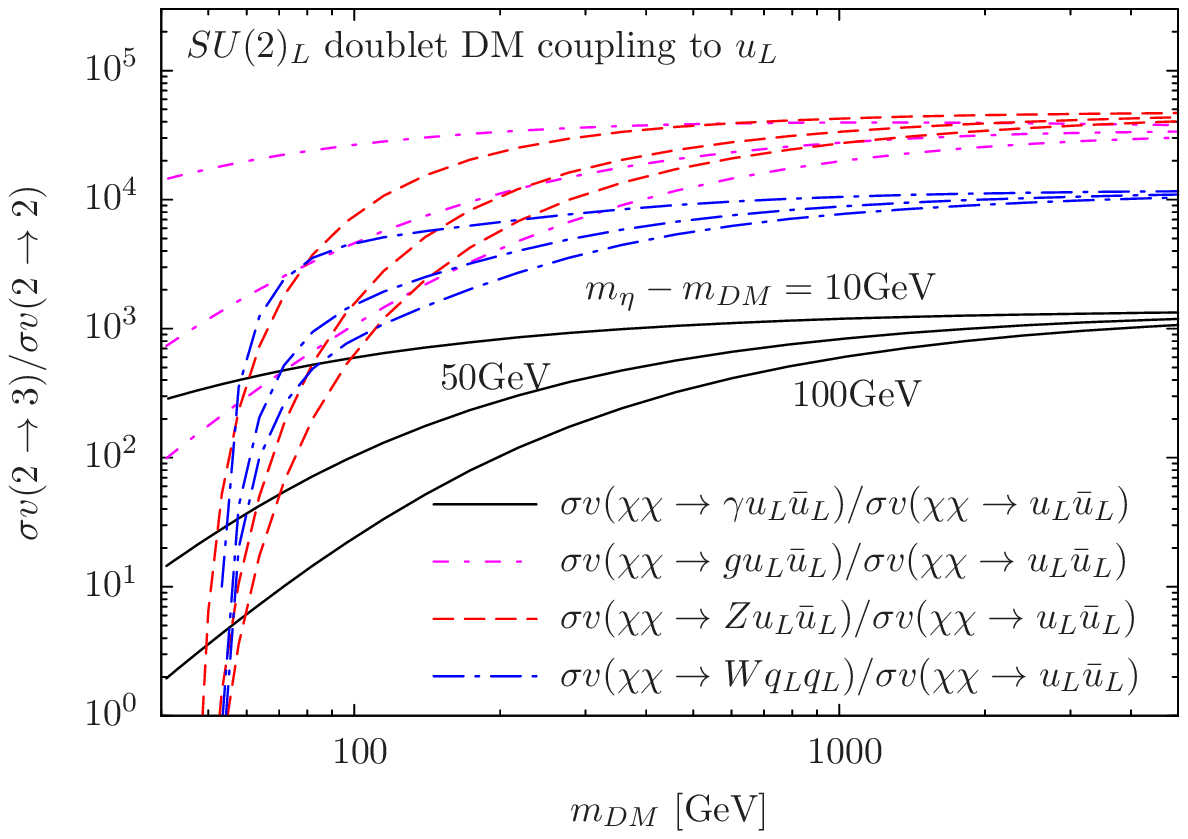}
 &\hspace*{-0.8cm} \includegraphics[width=0.55\textwidth]{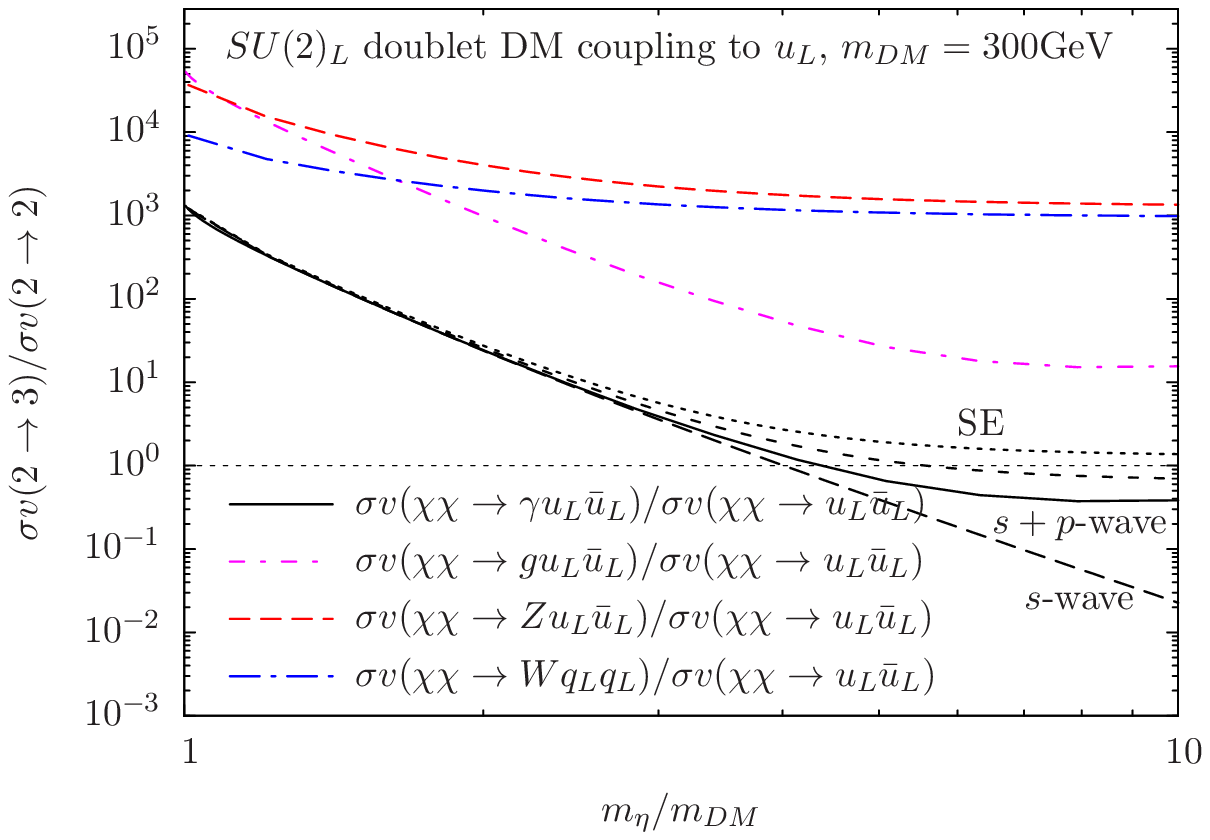} \\
\end{tabular}
 \caption{\label{fig:crossSection1_Doubet_Q} Ratio of three-body and two-body annihilation cross-sections. The top and bottom rows show the case of $SU(2)_L$ doublet dark matter coupling to right-handed up- and down-quarks with equals strength ($f=f'$), and to the left-handed up quark, respectively. The left column shows the dependence on the dark matter mass for fixed mass splitting $m_{\eta}-m_{\rm DM}=10, 50$ and $100$GeV, while the right column shows the dependence on the mass of the mediating scalar particle $\eta$ for $m_{\rm DM}=300$GeV. For the relative dark matter velocity we use $v=10^{-3}c$.}
\end{figure}

\begin{figure}
\begin{center}
 \includegraphics[width=0.95\textwidth]{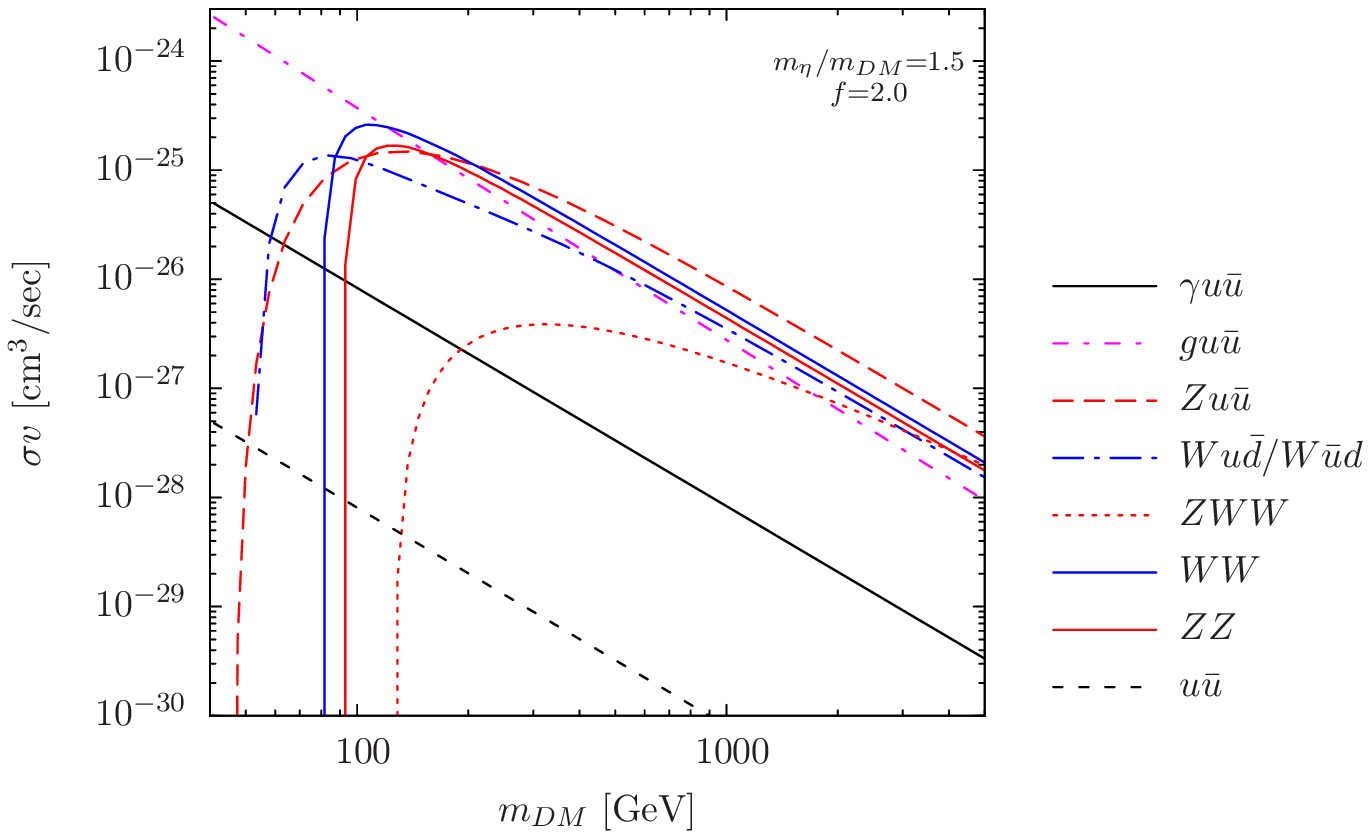} \\
 \includegraphics[width=0.95\textwidth]{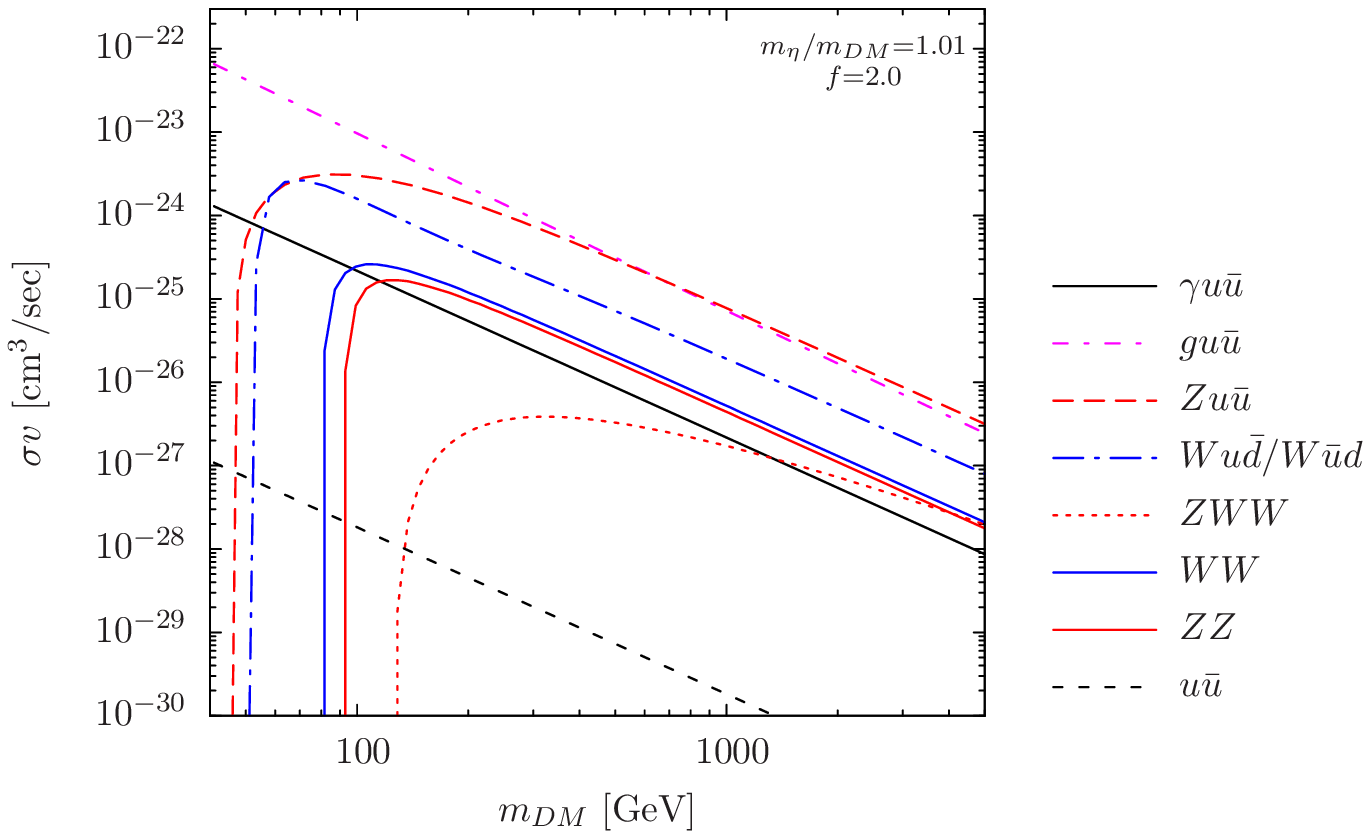}
\end{center}
 \caption{\label{fig:crossSection1_Doublet_UL_Gauge} Cross-sections of various annihilation channels for the case of
$SU(2)_L$ doublet dark matter coupling to the left-handed up quark, for $m_{\eta^\pm}/m_{\rm DM}=1.5$ (top)
and $1.01$ (bottom). In both cases, the Yukawa coupling is chosen as $f=2$. For the relative dark matter velocity we use $v=10^{-3}c$.}
\end{figure}

\subsection{$SU(2)_L$ doublet dark matter}

The case of doublet dark matter coupling to quarks can be discussed in close analogy to
the case with a coupling to leptons. In particular, the gauge quantum numbers of the
dark matter doublets $\chi_{1,2}$, their mass eigenstates and gauge interactions are
identical. Differences arise only with respect to the coupling to colored scalar particles $\eta$.
We will again discuss the different possibilities. The dark matter particle can couple to the
right-handed quarks via a mediating scalar $\eta\equiv (3,2,1/6)$,
\begin{align}
  \begin{split}
    {\cal L}^{\rm fermion}_{\rm int}&= f  (\bar\chi_1  i\sigma_2 \eta^*)d_R + f'  (\bar\chi_2  i\sigma_2 \eta^*)u_R+
    {\rm h.c.}\;.
  \end{split}
\end{align}
In contrast to the leptons, two coupling terms are allowed by the symmetries, coupling the dark matter
to the up- and down quarks, respectively. Alternatively, the dark matter particle can couple to the left-handed quark doublet. In this case, two choices for the quantum numbers of the mediating scalar field are possible. For $\eta\equiv(3,1,-1/3)$,
the interaction term reads
\begin{align}
  \begin{split}
    {\cal L}^{\rm fermion}_{\rm int}&= 
    f (\bar \chi_1 i\sigma_2 q_L^c) \eta+{\rm h.c.} = f\Big(\frac{1}{\sqrt{2}}(\bar\chi'+\epsilon\bar\chi)d_L^c-\bar\chi^\pm u_L^c\Big)+{\rm h.c.}\;,
  \end{split}
\end{align}
which couples the dark matter particle $\chi$ to the left-handed down quarks. The second possibility
is to choose $\eta\equiv(3,1,2/3)$. Then the interaction term reads
\begin{align}
  \begin{split}
    {\cal L}^{\rm fermion}_{\rm int}&= 
    f (\bar \chi_2 i\sigma_2 q_L^c) \eta+{\rm h.c.} = f\Big(-\frac{1}{\sqrt{2}}(\bar\chi'-\epsilon\bar\chi)u_L^c+\bar\chi^\pm d_L^c\Big)+{\rm h.c.}\;,
  \end{split}
\end{align}
and we obtain a coupling of dark matter to left-handed up quarks.
In a supersymmetric context, the scalars can be identified with $\tilde q_L$, $\tilde d_R$, and $\tilde u_R$, respectively.
These interactions will lead to annihilation into light quarks via internal bremsstrahlung of either a photon, a weak
gauge boson, or a gluon. For the photon and the gluon, the corresponding diagrams are identical to those obtained for singlet dark matter, and therefore the cross sections are also the same. However, electroweak bremsstrahlung can proceed also
via initial state radiation. As in the leptonic case, this leads to a considerable enhancement of the annihilation
into $Wq\bar q'$ and $Zq\bar q$. We will now discuss the branching ratios for each case.

\subsubsection*{Coupling to right-handed quarks}

The annihilation of dark matter into right-handed quarks can be mediated by a coloured scalar $\eta$
with quantum numbers $(3,2,1/6)$. There are two possible couplings, $f$ and $f'$, which correspond
to a coupling to $u_R$ and to $d_R$, respectively. Here, we assume for simplicity that both components
of $\eta$ are degenerate in mass.
The branching ratios in the limit $m_{\eta} \gg m_{\rm DM}\gg \frac{M_Z}{2}$ are given by
\begin{equation}
\begin{array}{ccccc}
  \sigma v(\chi\chi\to g q\bar q) & : & \sigma v(\chi\chi\to \gamma q\bar q) & \simeq & \frac{12(f^4+{f'}^4)}{4f^4+{f'}^4}\frac{\alpha_s(m_{\rm DM})}{\alpha_{em}}\;, \\
  \sigma v(\chi\chi\to Z q\bar q) & : & \sigma v(\chi\chi\to \gamma q\bar q) & \simeq & \frac{9}{(4f^4+{f'}^4)}\frac{f^4F(4s_W^2/3)+{f'}^4F(-2s_W^2/3)}{60 s_W^2c_W^2}\;,  \\
  \sigma v(\chi\chi\to W q\bar q') & : & \sigma v(\chi\chi\to \gamma q\bar q) & \simeq & \frac{9(ff')^2}{4f^4+{f'}^4}\frac{50\mu^2+12}{15 s_W^2}  \;,
\end{array}
\end{equation}
where $F(x)=50\mu(\mu+x)+15(1-x)^2-3$, $\mu=m_\eta^2/m_{\rm DM}^2$, and the cross sections denote the
sum of annihilation processes into up and down type quarks. As in the leptonic case, the ratio of cross sections
of electroweak to electromagnetic processes depends quadratically on $\mu$ because of the different scaling
of the cross sections with $\mu$ due to initial state radiation. In the upper right part of Fig.~\ref{fig:crossSection1_Doubet_Q} the dependence of the cross sections on the dark matter mass is shown for three
different mass splittings $m_\eta-m_{\rm DM}=10,50,100$GeV, assuming that $f=f'$. Note that the annihilation
into weak bosons can be as strong or even stronger than the annihilation into gluons. The reason is that
the former can proceed also via initial state radiation, which leads to a parametric enhancement compared
to the annihilation into gluons for large values of $\mu$. It turns out that even for moderate values of
$\mu$ the additional channels due to initial state radiation enhance the electroweak processes significantly.
Note, however, that the branching ratio into $W$ bosons gets suppressed when the ratio of couplings $|f/f'|$ deviates
from unity. The dependence of the branching ratios on $m_\eta$ is shown in the upper left part of Fig.~\ref{fig:crossSection1_Doubet_Q}. As expected, the cross sections for $gq\bar q$ and $\gamma q\bar q$ fall off
as $1/\mu^4\propto 1/m_\eta^8$, until the p-wave contribution dominates for $m_\eta\gtrsim 4m_{\rm DM}$. The
electroweak processes scale with $1/\mu^2$, because of the contribution from initial state radiation. Since also
the $2\to 2$ cross section scales like $1/\mu^2$, the ratio approaches a constant. In Fig.~\ref{fig:crossSection1_Doubet_Q}, also the leading effect of Sommerfeld enhancement is shown, for the same choice of parameters as
discussed in the leptonic case. As for a coupling to leptons, the main effect is to enhance the branching
fraction into $\gamma q\bar q$ due to the annihilation process $\chi\chi\to\chi^+\chi^-\to \gamma q\bar q$.
The corresponding cross section can be inferred from the formulae given in the Appendix.

\subsubsection*{Coupling to left-handed quarks}

There are two possibilities, $\eta$ can have the quantum numbers $(3,1,2/3)$ or $(3,1,-1/3)$.
In the first case it mediates a coupling of dark matter to $u_L$, and in the second to $d_L$.
Lets consider both possibilities separately. For $\eta=(3,1,2/3)$, the branching ratios in the limit $m_{\eta} \gg m_{\rm DM}\gg \frac{M_Z}{2}$ read
\begin{equation}
\begin{array}{ccccc}
  \sigma v(\chi\chi\to g u\bar u) & : & \sigma v(\chi\chi\to \gamma u\bar u) & \simeq & 3\alpha_s(m_{\rm DM})/\alpha_{em}\;, \\
  \sigma v(\chi\chi\to Z u\bar u) & : & \sigma v(\chi\chi\to \gamma u\bar u) & \simeq &  3F(1-4s_W^2/3)/(80 s_W^2c_W^2)\;,  \\
  \sigma v(\chi\chi\to W q\bar q') & : & \sigma v(\chi\chi\to \gamma u\bar u) & \simeq & 3(50\mu(\mu-1)+63)/(80s_W^2)  \,.
\end{array}
\end{equation}
For $\eta=(3,1,-1/3)$, on the other hand, the branching ratios are given by
\begin{equation}
\begin{array}{ccccc}
  \sigma v(\chi\chi\to g d\bar d) & : & \sigma v(\chi\chi\to \gamma d\bar d) & \simeq & 12\alpha_s(m_{\rm DM})/\alpha_{em}\;, \\
  \sigma v(\chi\chi\to Z d\bar d) & : & \sigma v(\chi\chi\to \gamma d\bar d) & \simeq &  3F(2s_W^2/3-1)/(20 s_W^2c_W^2)\;,  \\
  \sigma v(\chi\chi\to W q\bar q') & : & \sigma v(\chi\chi\to \gamma d\bar d) & \simeq & 3(50\mu(\mu-1)+63)/(20s_W^2)  \;,
\end{array}
\end{equation}
where the function $F(x)$ is the same as defined above. The dependence on the
dark matter mass and on the mass of the mediating scalar is shown in the lower part of
Fig.~\ref{fig:crossSection1_Doubet_Q} for the case $\eta=(3,1,2/3)$. The behaviour is qualitatively similar to
the one discussed above. In addition, the absolute values of the various cross sections
are shown in Fig.~\ref{fig:crossSection1_Doublet_UL_Gauge} for the same choice of parameters than for
the case of a coupling to leptons. Due to the color enhancement and the annihilation channel into gluons,
the annihilation processes mediated by the scalars $\eta$ are important over the whole dark matter mass
range for both choices $m_\eta/m_{\rm DM}=1.5$ and $1.01$.

\section{Antiproton-to-proton ratio and observational constraints}\label{sec:pbar}

\begin{figure}
\includegraphics{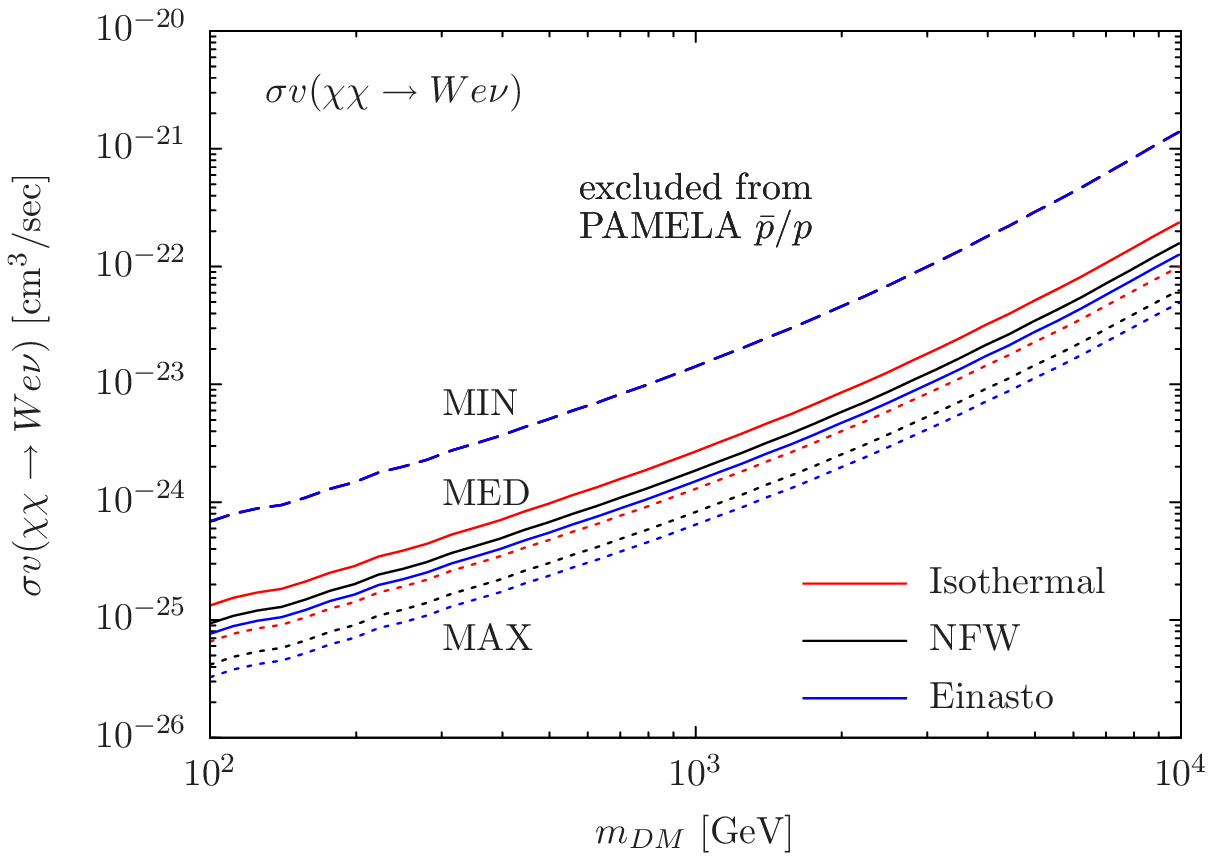}
\includegraphics{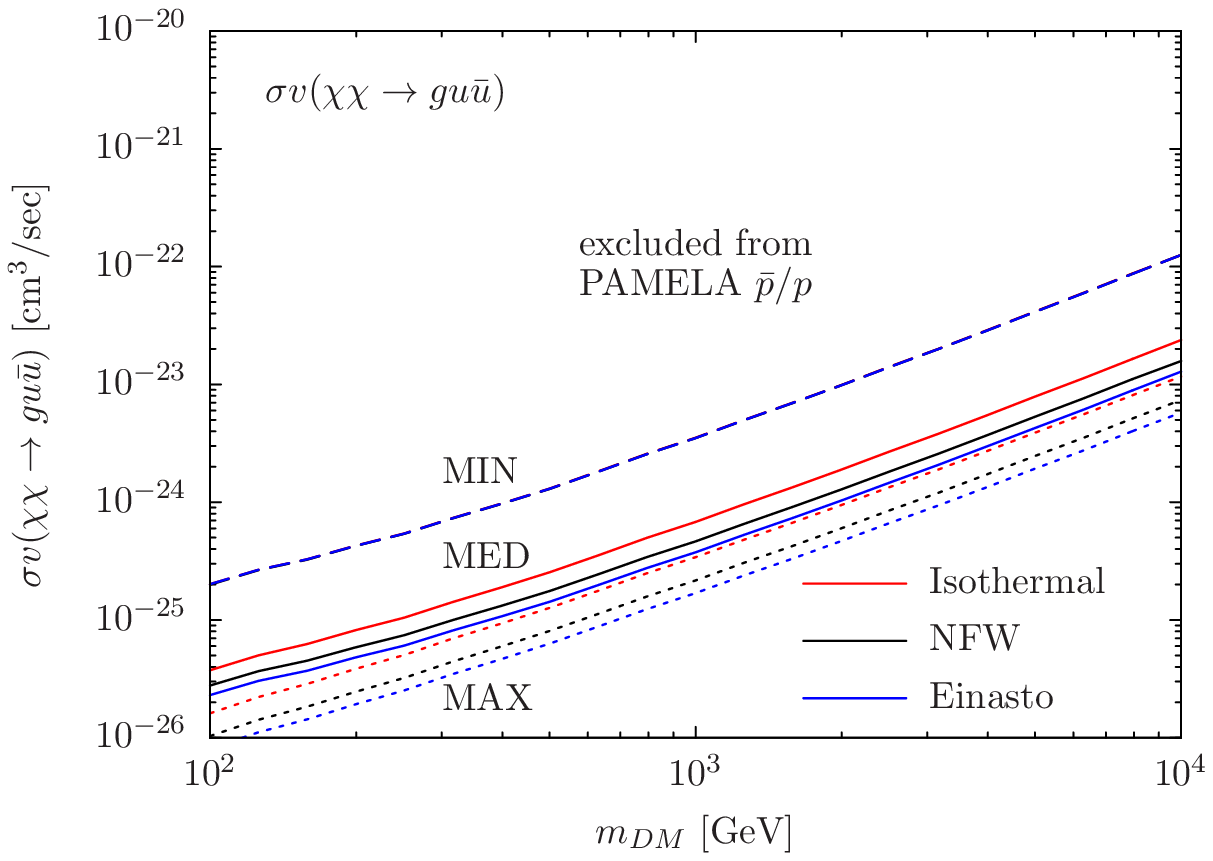}
 \caption{\label{fig:exclusionPlot1} Upper bounds on the electroweak IB cross-sections $\sigma v(2\to 3)$ obtained from the PAMELA data \cite{Adriani:2010rc} of the cosmic antiproton-to-proton ratio. The plots correspond to the constraints on the individual electroweak IB processes $\chi\chi\to W e\nu$ (where $W e\nu\equiv W^- \bar e\nu + W^+ e\bar \nu$, upper), and $\chi\chi\to g u\bar u$ (lower) at 95\%C.L. Dashed lines correspond to the MIN, solid to MED and dotted to MAX propagation models. The bounds obtained assuming an isothermal dark matter profile are shown in red, NFW in black, and Einasto in blue.}
\end{figure}

Two dark matter particles at the position $\vec r$ can annihilate 
producing antiprotons at a rate per unit of kinetic energy and volume
given by:
\begin{align}
Q(T,\vec r)=\frac{1}{2}\frac{\rho^2(\vec r)}{m^2_{\rm DM}}
\sum_f \langle \sigma v\rangle_f \frac{dN^f_{\bar p}}{dT}\;,
\label{eq:source}
\end{align}
where $\langle \sigma v\rangle_f$ is the thermally averaged cross-section
multiplied by the velocity in the annihilation channel $f$, $\rho(\vec r)$ is the
distribution of dark matter particles in the Milky Way, where $\vec r$ denotes
the position of the dark matter particle with respect to the center
of our Galaxy, and $dN^f_{\bar p}/dT$ is the energy spectrum of antiprotons produced in 
that channel per unit of kinetic energy. We will assume for simplicity
a spherically symmetric distribution, and calculate the antiproton flux assuming
a radial dependence given by either the Isothermal, NFW or Einasto profile with parameters
specified in \cite{Garny:2011cj}. The spectrum of antiprotons is obtained using the event generator PYTHIA 8.1 \cite{Sjostrand:2007gs} interfaced with CALCHEP \cite{Pukhov:1999gg, Pukhov:2004ca}.

After being produced at the position $\vec r$, antiprotons propagate
through the Milky Way in a complicated way before reaching the Earth.
Following \cite{ACR}, we will describe antiproton propagation by means of
a stationary two-zone diffusion model with cylindrical boundary conditions.
Under this approximation, the number density of antiprotons
per unit kinetic energy, $f_{\bar p}(T,\vec{r},t)$, approximately 
satisfies the following transport equation:
\begin{equation}
0=\frac{\partial f_{\bar p}}{\partial t}=
\nabla \cdot (K(T,\vec{r})\nabla f_{\bar p})
-\nabla \cdot (\vec{V_c}(\vec{r})  f_{\bar p})
-2 h \delta(z) \Gamma_{\rm ann} f_{\bar p}+Q(T,\vec{r})\;.
\label{transport-antip}
\end{equation}
The boundary conditions require the solution $f_{\bar p}(T,\vec{r},t)$ to vanish at the boundary of the diffusion zone, which is approximated by a cylinder with half-height $L = 1-15~\rm{kpc}$ and radius $ R = 20 ~\rm{kpc}$. The diffusion in the Galactic magnetic field and the convection term, which accounts for the drift of charged particles away from the disk induced by the Milky Way's Galactic wind, are described by the parameterization $K(T)=K_0 \;\beta\; {\cal R}^\delta$ and $\vec{V}_c(\vec{r})=V_c\; {\rm sign}(z)\; \vec{k}$. The third term accounts for antimatter annihilation with rate $\Gamma_{\rm ann}$, when it interacts with ordinary matter in the Galactic disk. In order to take the uncertainties related to propagation into account, we will use three sets of parameters, compatible with the cosmic boron to carbon flux ratio~\cite{Maurin:2001sj}, corresponding to minimum, medium and maximum antiproton flux as given in Table \ref{tab:param-antiproton}. The flux at the position of the solar system is given by $\Phi^{\rm{IS}}_{\bar p}(T) = \frac{v}{4 \pi} f_{\bar p}(T,r_\odot)$. Finally, we take the effect of solar modulation into account using the force field approximation~\cite{solar-modulation,perko} with solar modulation parameter $\phi_F=500$ MV for our numerical analysis. In order to obtain constraints on the dark matter annihilation cross-section, we compute the antiproton-to-proton ratio $ \bar p/p \equiv (\Phi_{\bar p}^{\rm sig} +  \Phi_{\bar p}^{\rm bkg} )/\Phi_p $ using the proton flux of \cite{Bringmann:2006im}. The background flux arises from  antiprotons that are produced by spallation of cosmic ray nuclei, mainly protons and Helium, on the interstellar medium. We use the backgound flux calculated in Ref. \cite{Donato:2001ms} based on the two-zone diffusion model, taking into account p-p, He-p, p-He and He-He nuclear reactions. The main uncertainties arise from the diffusion parameters and the nuclear cross-sections, and are estimated to be in the range of $10-25$\% depending on the energy. In contrast, the uncertainty stemming from the knowledge of the primary flux of cosmic nuclei and the composition of the interstellar medium are found to be subdominant. Note that the prediction for the secondary antiproton flux is consistent with the PAMELA data  \cite{Adriani:2010rc}. In order to obtain a conservative exclusion bound we adopt the minimal value for the antiproton background as discussed in \cite{Donato:2001ms}. Upper limits on the cross-sections of the individual annihilation channels, as well as on astrophysical boost factors, are then obtained from the PAMELA $\bar p/p$ data \cite{Adriani:2010rc} using a $\chi^2$-test at $95\%$C.L.

\begin{table}[t]
\begin{center}
\begin{tabular}{|c|cccc|}
\hline
Model & $\delta$ & $K_0\,({\rm kpc}^2/{\rm Myr})$ & $L\,({\rm kpc})$
& $V_c\,({\rm km}/{\rm s})$ \\
\hline 
MIN & 0.85 & 0.0016 & 1 & 13.5 \\
MED & 0.70 & 0.0112 & 4 & 12 \\
MAX & 0.46 & 0.0765 & 15 & 5 \\
\hline
\end{tabular}
\caption{\label{tab:param-antiproton} Astrophysical parameters compatible with the B/C ratio that yield the minimal (MIN), median (MED) and maximal (MAX) flux of antiprotons.}
\end{center}
\end{table}

\begin{figure}
\includegraphics{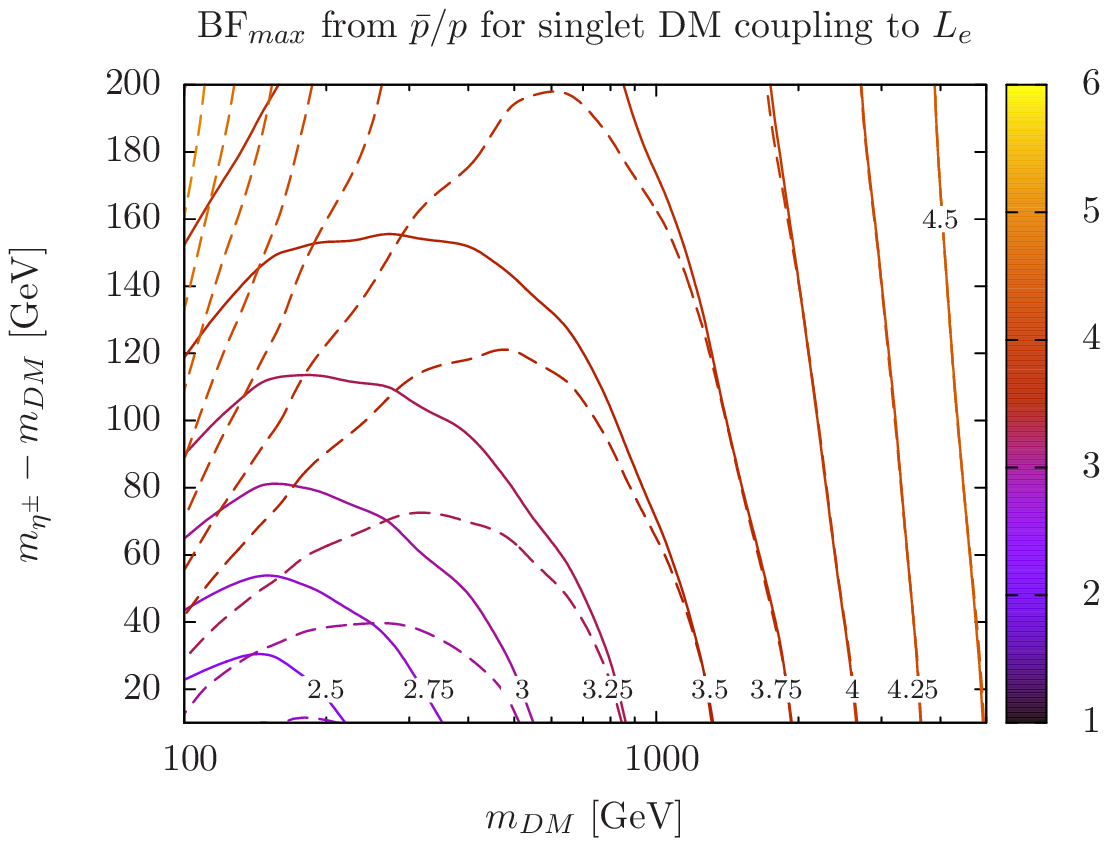}
\includegraphics{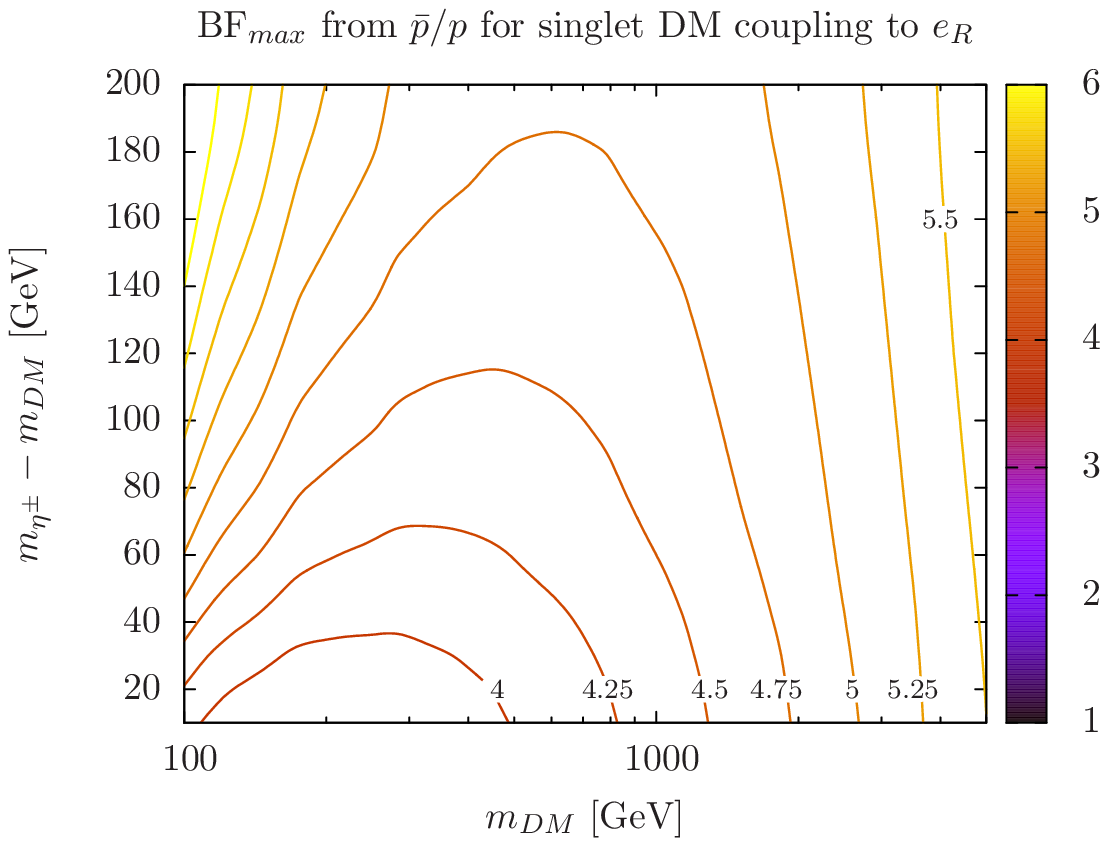}
 \caption{\label{fig:exclusionPlot_EL} Upper bounds on the astrophysical boost factor for singlet dark matter coupling
to leptons. Here we use the MED propagation parameters and the NFW profile. The contours correspond to the maximally allowed
boost factor $\log_{10}(BF)$ from the PAMELA $\bar p/p$ data \cite{Adriani:2010rc}. The coupling $f$ is fixed by requiring that the thermal relic density matches the WMAP
value. The upper plot shows the case of annihilations into left-handed leptons. The solid lines correspond to $m_{\eta^0}^2-m_{\eta^\pm}^2=v_{EW}^2$ and the
dashed lines to $m_{\eta^0}=m_{\eta^\pm}$. In the lower plot the constraints for dark matter annihilating to right-handed
electrons are shown.}
\end{figure}

\begin{figure}
\includegraphics{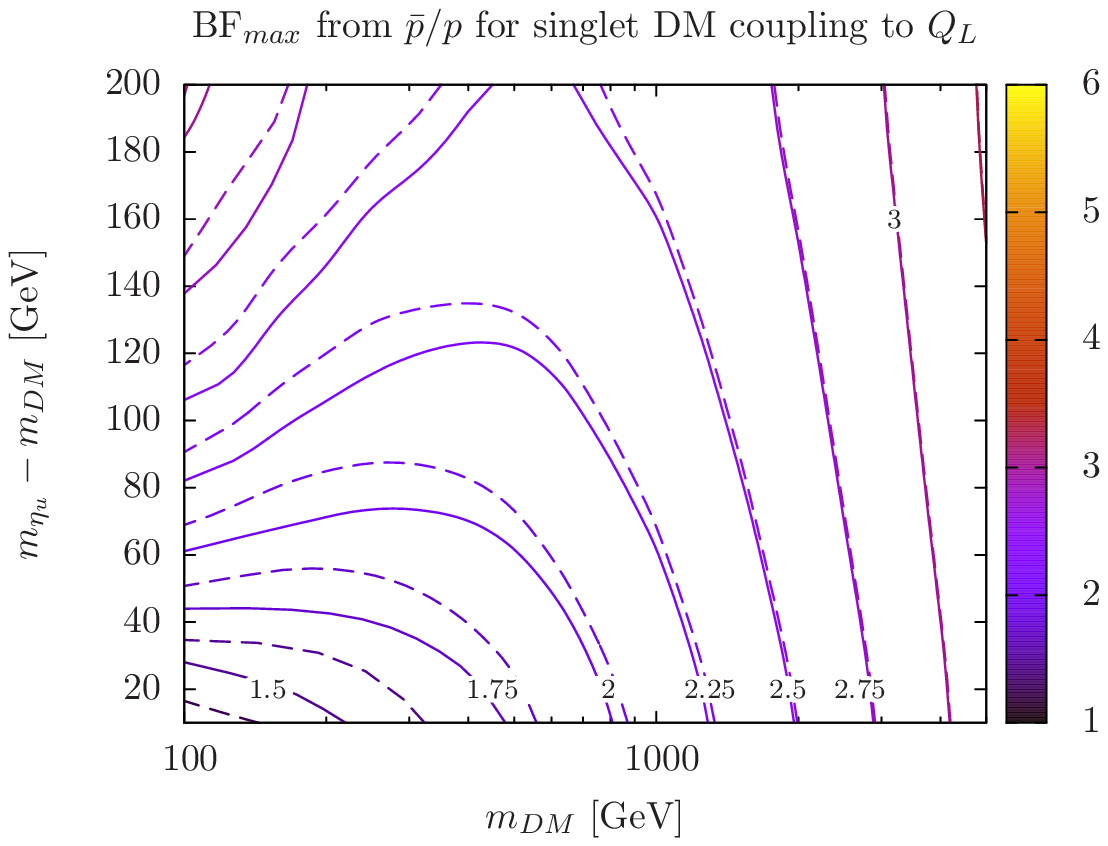}
\includegraphics{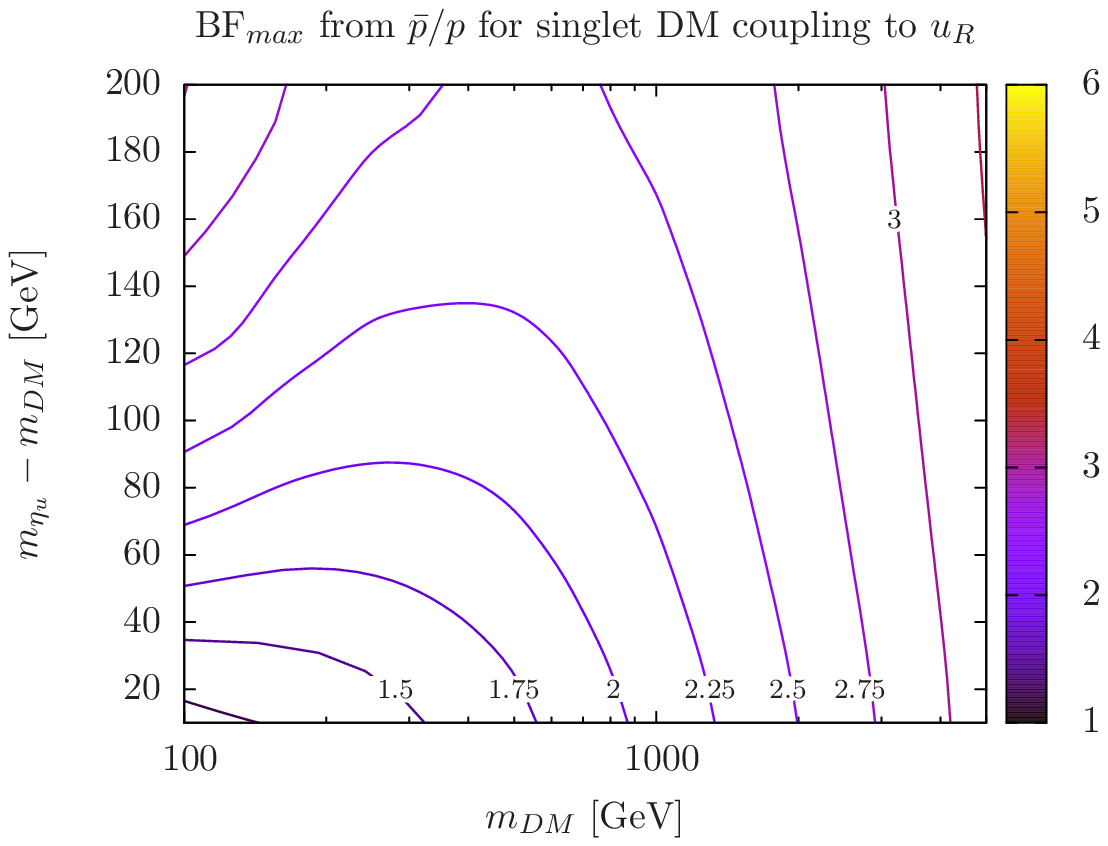}
 \caption{\label{fig:exclusionPlot_QL} Upper bounds on the astrophysical boost factor for singlet dark matter coupling
to quarks, obtained under the same assumptions as described in Fig.~\ref{fig:exclusionPlot_EL}. The upper plot shows the case of annihilations into left-handed up and down quarks. The solid lines correspond to $m_{\eta_u}^2-m_{\eta_d}^2=v_{EW}^2$ and the
dashed lines to $m_{\eta_u}=m_{\eta_d}$. In the lower plot the constraints for dark matter annihilating to right-handed
up quarks are shown. The relevant IB process is $\chi\chi\to gq\bar q$.}
\end{figure}

\begin{figure}
\includegraphics{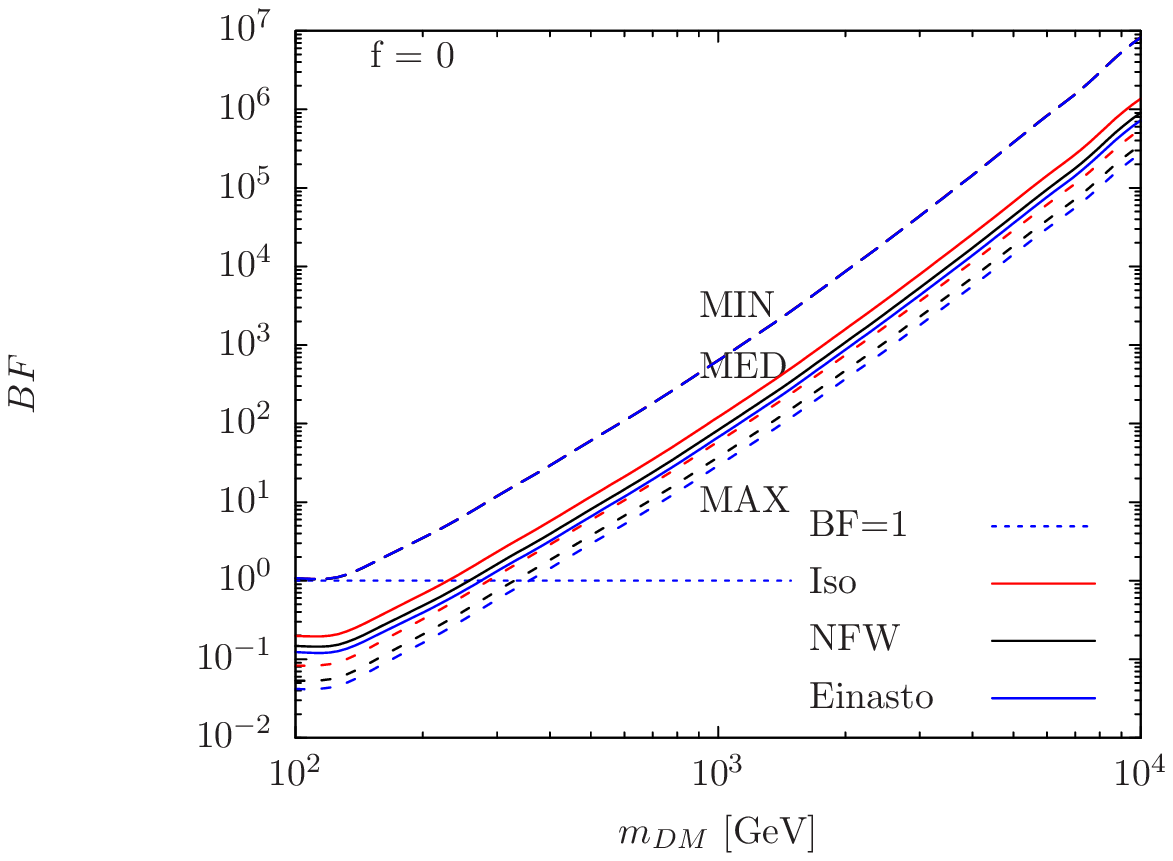}
\includegraphics{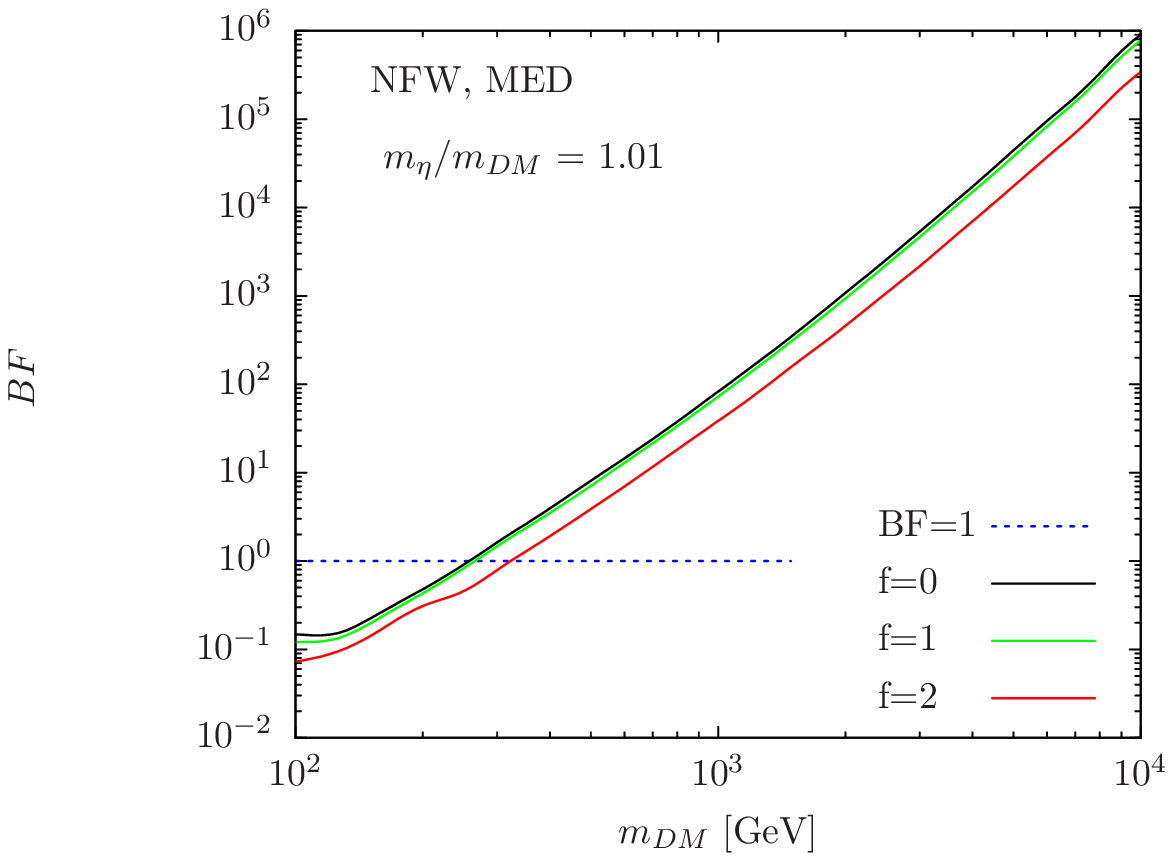}
 \caption{\label{fig:higgsino-like} Upper bounds on the astrophysical boost factor for doublet dark matter coupling to gauge bosons and leptons. The upper plot shows the case of dark matter annihilating exclusively into gauge bosons. In the lower plot we compare the bounds that arise from including leptonic final states with a coupling $f=1$ and $ f=2$ to the case without leptons. For a better comparison the astrophysical parameters have been fixed to a NFW profile and MED propagation in the lower plot.}
\end{figure}

We show in Fig.\,\ref{fig:exclusionPlot1} our results for the upper limits on the cross-sections of individual IB processes. In order to take the astrophysical uncertainties into account, we compute limits for the three propagation models and three halo profiles discussed above. The limits on the individual channels do not depend significantly on the mass spectrum as long as $m_{\eta^i}/m_{\rm DM}\sim\mathcal{O}(1)$. For the MED propagation model the limits lie in the range $\sigma v(\chi \chi\rightarrow W e\nu)\lesssim 10^{-25}\cm^3/\sec$ for $m_{\rm DM}=100\GeV$ and $10^{-24}\cm^3/\sec$ for $m_{\rm DM}=1\TeV$ for electroweak IB processes involving leptons. For dark matter coupling to quarks the IB of gluons is the dominant source of antiprotons. The limit lies in the range $\sigma v(\chi \chi\rightarrow g u\bar u)\lesssim 3\times 10^{-26}\cm^3/\sec$ for $m_{\rm DM}=100\GeV$ and $5\times 10^{-25}\cm^3/\sec$ for $m_{\rm DM}=1\TeV$. These limits can be compared to the ones recently derived in \cite{Bringmann:2012vr} on the annihilation cross section of singlet dark matter particles into right-handed fermions and a photon. The non-observation by the Fermi-LAT of an excess of gamma-rays translates into the limits  $\sigma v(\chi \chi\rightarrow \gamma \mu^+ \mu^-)\lesssim 6\times 10^{-28}\cm^3/\sec$, $\sigma v(\chi \chi\rightarrow \gamma \tau^+ \tau^-)\lesssim 6\times 10^{-28}\cm^3/\sec$, $\sigma v(\chi \chi\rightarrow \gamma b\bar b)\lesssim 4\times 10^{-28}\cm^3/\sec$ for $m_{\rm DM}=100\GeV$. For dark matter coupling to leptons, the upper limit on the cross section on the process $\chi \chi\rightarrow W e\nu$ from the PAMELA antiproton data roughly translates, using  Eq.~\ref{eq:ratios-lepto}, into  $\sigma v(\chi \chi\rightarrow \gamma \ell^+ \ell^-)\lesssim 2\times 10^{-26}\cm^3/\sec$, which is weaker than the limit obtained in \cite{Bringmann:2012vr} from the Fermi-LAT data. For larger masses, where there are no limits from gamma-rays, the PAMELA data provide the strongest constraints on the model. On the other hand, for dark matter coupling to quarks, one obtains from the PAMELA antiproton data and Eqs.~(\ref{eq:ratios-uR},~\ref{eq:ratios-dR},~\ref{eq:ratios-qL}) the upper limits $\sigma v(\chi \chi\rightarrow \gamma q\bar q)\lesssim 2\times 10^{-28}\cm^3/\sec,~7\times 10^{-28} \cm^3/\sec,~5\times 10^{-28}\cm^3/\sec$  for couplings to down-type quark singlets, up-type quark singlets and quark doublets, respectively, and $m_{\rm DM}=100\GeV$. Hence, in this range of dark matter masses the antiproton constraints on the model can be, depending on the propagation model, competitive with the gamma-ray constraints.

The limits on the cross-sections can be translated into limits on the coupling strength of dark matter to Standard Model particles within the scenarios discussed in the previous sections. In turn, these can be used to compute limits on an astrophysical boost factor. For the case of singlet dark matter, these limits are computed by fixing the coupling $f$ such that thermal freeze out produces a relic density in agreement with the WMAP value $\Omega_\chi h^2 \simeq 0.11$. The required value of the coupling parameter $f$ depends on the mass spectrum and can be found in Ref.~\cite{Cao:2009yy,Garny:2011cj}. For the case of doublet dark matter, the gauge coupling is fixed and therefore the above prescription would be too restrictive. Therefore, we do not impose the relic density constraint in that case. Instead, we compute the antiproton flux arising from the annihilations of the dark matter particles in the galaxy using the physical value of the gauge coupling constant and fixed values $f=0,1,2$ of the Yukawa coupling. A limit on the boost factor can then be obtained by requiring that the antiproton flux obtained from the annihilations of doublet dark matter for a given set of masses and coupling constants, and multiplied by the boost factor, is consistent with the PAMELA $\bar p/p$ data at the 95\%C.L. Note that the maximally allowed boost factor determined in this way can be formally less than unity, which means that the corresponding model parameters can be excluded. (old: Instead, we compute the boost factor for the case of doublet dark matter by keeping the ratio of Yukawa- and gauge coupling at a fixed value, and rescale all the fluxes according to a common boost factor.)

The results for singlet dark matter coupling to leptons are shown in Fig.~\ref{fig:exclusionPlot_EL},
in dependence on the dark matter mass and the mass splitting between the scalar particle $\eta$ mediating
the annihilation and the dark matter particle. In Fig.~\ref{fig:exclusionPlot_QL}, the corresponding
constraints for singlet dark matter coupling to quarks are shown. For leptons, the relevant channels are $\chi\chi\to We\nu$, $\chi\chi\to Ze\bar e$
and $\chi\chi\to Z\nu\bar \nu$. For quarks, the antiproton flux is dominantly produced by the
process $\chi\chi\to g q\bar q$. As expected, the constraints are significantly stronger than for leptons,
mainly because the internal bremsstrahlung of gluons has a larger cross section and the colored
particles in the final state lead to a more efficient antiproton production.

In Fig.~\ref{fig:higgsino-like} the results for doublet dark matter coupling to leptons are shown. In this case the channels $ \chi \chi \to W^+ W^-$ and  $\chi \chi \to Z Z$ are relevant for most couplings and mass splittings. Only for a coupling $f > 1$ and a mass splitting $m_{\eta}/m_{\rm DM} \approx 1.01 $ non-negligible contributions arise from the leptonic channels $\chi \chi \to W e \nu$ and $\chi\chi\to Ze\bar e$. 
It should be noted that doublet dark matter with masses $m_{\rm DM} \lesssim 200 \GeV$ can be excluded due to the antiproton constraints even without any additional antiprotons produced in leptonic channels. This reach can be extended through the inclusion of leptonic channels. As was to be expected the constraints worsen significantly for dark matter masses $m_{\chi} > 1 \TeV$, so that doublet dark matter is virtually unconstrained by antiprotons in the $\TeV$ range.

In general, the antiproton constraints are most stringent for small mass splittings $m_\eta-m_{\rm DM}$, because the internal bremsstrahlung cross sections are strongly enhanced. Note that in this parameter region the bounds from collider searches are weakest, as discussed before.
 
\section{Conclusions}\label{sec:conclusions}

We have studied the annihilation process of two Majorana dark matter particles $\chi\chi\rightarrow f\bar f V$, with $f$ a light Standard Model fermion and $V$ a gauge boson, which is relevant for their indirect detection. Under very general conditions, this process dominates over the lowest order annihilation process $\chi\chi\rightarrow f\bar f$, due to the lifting of the helicity supression in the $s$-wave contribution to the cross section of the $2\rightarrow 2$ process  by the associated emission of a spin 1 particle.

To keep the analysis as general as possible, we have performed an extensive analysis of possible dark matter scenarios, focusing on the plausible case where the dark matter particle is a $SU(2)_L$ singlet or doublet. We have classifed all scenarios where the dark matter particle couples to the first generation of Standard Model fermions and an extra scalar particle, which mediates the annihilations  $\chi\chi\rightarrow f\bar f$ and $\chi\chi\rightarrow f\bar f V$. The gauge invariance of the Yukawa coupling requires the intermediate scalar particle to carry hypercharge and electric charge. Hence, the helicity suppression of the $2\rightarrow 2$ process can be lifted through the associated emission of a photon or a weak gauge boson off the final charged fermions or off the intermediate charged scalar, which is more efficient when the mass of the intermediate scalar is close to the mass of the dark matter particle. In the case of the $SU(2)_L$ doublet, the weak gauge boson can also be emitted off the initial dark matter particle, resulting in an enhancement of the cross section with respect to the $SU(2)_L$ singlet case. Moreover, when the two weak isospin components of the $SU(2)_L$ doublets have a sizable mass splitting, the annihilation cross section into weak gauge bosons is further enhanced by the radiation of longitudinally polarized $W$-bosons. We have provided analytic expressions for the cross sections for all these processes, complementing results already existing in the literature, and we have studied numerically the relative strength of each of the annihilation channels. If the dark matter particle couples to a light quark via a Yukawa coupling with a colored scalar, then not only the radiation of a photon or a weak gauge boson can lift the helicity suppression, but also the radiation of a gluon.

Models with exotic charged or colored particles are strongly constrainted by the negative searches at accelerators of an excess over the Standard Model expectations of dilepton or dijet events with missing energy. We have remarked that the choices of parameters where the $2\rightarrow 3$ processes are most important, namely when the intermediate scalar mass is close to the dark matter mass, evade the stringent lower bounds on the masses of the exotic particles from accelerator searches, since the lepton or the jet produced in the decay of the exotic charged or colored particle is too soft to pass all the cuts required by the current analyses.

Lastly, we have calculated the constraints on the various dark matter scenarios from the non-observation of an excess in the cosmic antiproton-to-proton fraction measured by the PAMELA collaboration. We have presented these constraints as upper limits on the cross section of the relevant annihilation processes, and translated them into constraints on an astrophysical boost factor for each scenario and a broad range of dark matter masses as well as mass splittings between dark matter
and the scalar particle mediating the annihilation.

\section*{Acknowledgements}

We thank F.~Calore for pointing out a typo in Eq.~(\ref{Wffprime}). The work of AI and SV was partially supported by the DFG cluster of excellence ``Origin and Structure of the Universe.'' SV acknowledges support from the DFG Graduiertenkolleg ``Particle Physics at the Energy Frontier of New Phenomena.''

\section*{Note added}

During the last stages of this work we became aware of another group discussing the annihilation of Majorana dark matter particles into quarks and a gluon~\cite{Asano:2011ik}.

\begin{appendix}

\section{Cross sections}

\subsection{$SU(2)_L$ singlet dark matter}

The differential cross-sections for the two-to-three dark matter annihilation processes $\chi\chi\to V f\bar f$
in the limit $v,m_f\to 0$, for the case of $SU(2)_L$-singlet Majorana dark matter $\chi$ coupling to the SM fermions $f$ via a mediating scalar $\eta_f$ are given by
\begin{eqnarray}
\frac{vd\sigma(\chi\chi\to\gamma f\bar f)}{dE_\gamma dE_f} & = & \frac{C_{\gamma f \bar f}\alpha_{em}f^4(1-x)[x^2-2x(1-y)+2(1-y)^2]}{8\pi^2 m_{\rm DM}^4 (1-2y-\mu_f)^2(3-2x-2y+\mu_f)^2} \\
\frac{vd\sigma(\chi\chi\to Z f\bar f)}{dE_Z dE_f} & = & \frac{C_{Z f \bar f}\alpha_{em}f^4 }{8\pi^2 m_{\rm DM}^4 (1-2y-\mu_f)^2(3-2x-2y+\mu_f)^2} \;, \nonumber \\
&& {} \times \Big\{ (1-x)[x^2-2x(1-y)+2(1-y)^2] \nonumber \\
&& {} + x_0^2[x^2+2y^2+2xy-4y]/4 -x_0^4/8 \Big\} \;, \\
\frac{vd\sigma(\chi\chi\to W f\bar f')}{dE_W dE_f} & = & \frac{C_{W f \bar f'} \alpha_{em}f^4 }{8\pi^2 m_{\rm DM}^4 (1-2y-\mu_f)^2(3-2x-2y+\mu_{f'})^2} \nonumber \\
&& {} \times \Big\{ (1-x)[x^2-2x(1-y)+2(1-y)^2+2(2-x-2y)\Delta\mu] \nonumber \\  
&& {} + x_0^2[x^2+2y^2+2xy-4y+2(2-x-2y)\Delta\mu+\Delta\mu^2]/4 -x_0^4/8  \nonumber\\
&& {} +\Delta\mu^2 [ (1-2x)/2 - 2(1-y)(1-x-y)/x_0^2 ]\Big\} \;, \label{Wffprime} \\
\frac{vd\sigma(\chi\chi\to g f\bar f)}{dE_\gamma dE_f} & = & \frac{C_{g f \bar f}\alpha_{s}(m_{\rm DM})f^4(1-x)[x^2-2x(1-y)+2(1-y)^2]}{8\pi^2 m_{\rm DM}^4 (1-2y-\mu_f)^2(3-2x-2y+\mu_f)^2} \;.
\end{eqnarray}
Here, $x=E_V/m_{\rm DM}$ for $V=\gamma,W,Z,g$, $y=E_f/m_{\rm DM}$, $x_0=M_V/m_{\rm DM}$, $\mu_f=m_{\eta_f}^2/m_{\rm DM}^2$,
$\mu_{f'}=m_{\eta_{f'}}^2/m_{\rm DM}^2$, and $\Delta\mu=(\mu_{f'}-\mu_f)/2$. The pre-factors are given by the following
expressions
\begin{center}
\begin{tabular}{c|cccc}
& $C_{\gamma f \bar f}$ & $C_{Z f \bar f}$ & $C_{W f \bar f'}$ & $C_{g q \bar q}$\\
\hline
$\chi\chi\to Vf_R\bar f_R$ & $q_f^2N_c$ & $q_f^2N_c\tan^2(\theta_W)$ & -- & $N_cC_F$ \\
$\chi\chi\to Vf_L\bar f_L$ & $q_f^2N_c$ & $\frac{(t_{3f}-q_f\sin^2(\theta_W))^2}{\sin^2(\theta_W)\cos^2(\theta_W)}N_c$ & $\frac{N_c}{2\sin^2(\theta_W)}$ & $N_cC_F$
\end{tabular}
\end{center}
where $q_f$ and $t_{3f}$ are the electric charge and the weak isospin, respectively, with $q_e=-1$ and $t_{3e}=-1/2$.
For quarks, one has $N_c=3$, and $C_F=(N_c^2-1)/(2N_c)=4/3$. The spectra of the vector bosons can be
obtained by integrating the differential cross-section over the fermion energy, with integration limits given by $E_f^{min/max}=m_{\rm DM}-(E_V\pm\sqrt{E_V^2-M_V^2})/2$. The total cross-section can be obtained by integrating over the remaining energy with limits $E_V^{min}=M_V$ and $E_V^{max}=m_{\rm DM}+M_V^2/(4m_{\rm DM})$. 
The relevant Feynman diagrams are shown in part (a) and (b) of Fig.~\ref{fig:Z_Diagram} for the exemplary case $\chi\chi\to Ze\bar e$.

For comparison, we also quote the leading contribution to the two-to-two cross section for $v\to 0$, in the
limit $m_f=0$,
\begin{equation}
\sigma v(\chi\chi\to  f\bar f) = \frac{N_cf^4v^2}{48\pi m_{\rm DM}^2}\frac{1+\mu_f^2}{(1+\mu_f)^4} \;,
\end{equation}
where $v$ is the relative velocity.

\subsection{$SU(2)_L$ doublet dark matter}

Here we report the $2\to 3$ cross sections for the case of $SU(2)_L$-doublet Majorana dark matter $\chi$ that
arise from a coupling to the SM fermions $f$ via a mediating scalar $\eta_f$. There can be additional contributions
due to $2\to 2$ annihilations into $WW$ or $ZZ$, with a subsequent decay of one of the gauge bosons. We do not include them
here for simplicity. Their size depends on the ratio $g/f$ of gauge interactions
and the interactions with the scalar $\eta_f$.

\begin{figure}
 \begin{center}
  \includegraphics{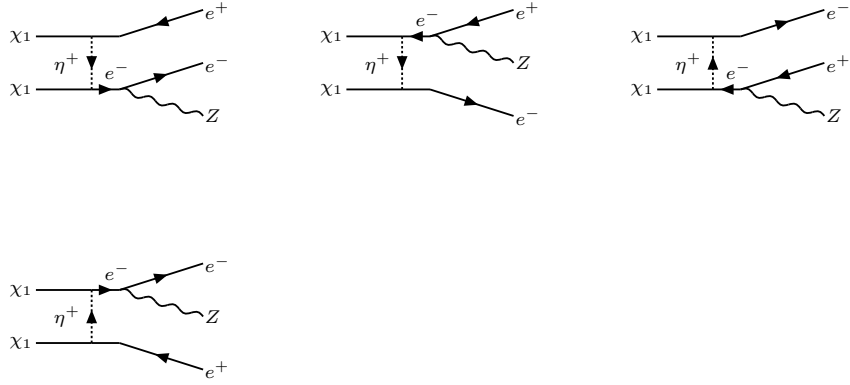}\\[1ex](a)\\[3ex]
  \includegraphics{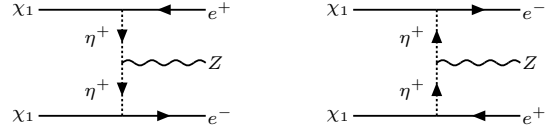}\\[1ex](b)\\[3ex]
  \includegraphics{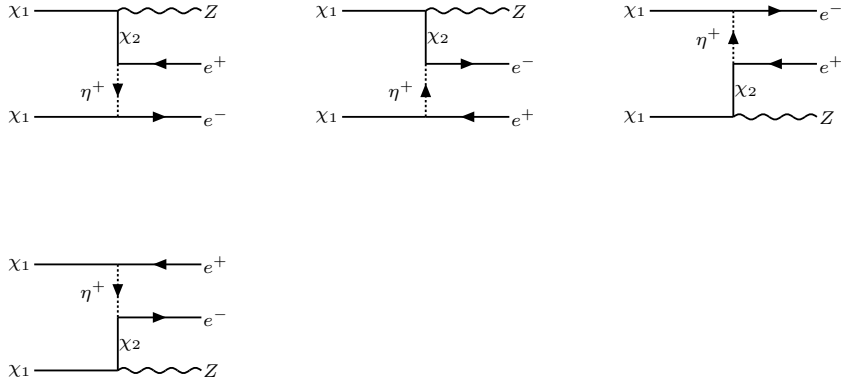}\\[1ex](c)
 \end{center}
 \caption{\label{fig:Z_Diagram} Feynman diagrams contributing to the annihilation channel $\chi\chi\to Ze\bar e$. For a singlet dark matter particle $\chi$ the diagrams corresponding to final state radiation (a) and virtual internal bremsstrahlung (b) contribute. Note that only their sum is gauge invariant. For a doublet dark matter particle $\chi$, also the diagrams (c) where the $Z$-boson is emitted from the initial state have to be taken into account in addition to (a) and (b). Here $\chi_1$ and $\chi_2$ refer to the two neutral mass eigenstates as explained in section \ref{doublet}. Their mass splitting is assumed to be negligibly small in Eq.\,(\ref{doubletZff}).}
\end{figure}

The main difference compared to the case of $SU(2)_L$ singlet dark matter are additional contributions from initial
state radiation. For annihilation into $\gamma f\bar f$ or $g f \bar f$, there are no such contributions because the dark matter is
electrically neutral and uncoloured. Therefore, the cross-section is identical to the one from above. The cross-section for
the annihilation to $Z f\bar f$ is given by a sum of ten diagrams: four with initial and final state radiation,
respectively, and two where the $Z$ is emitted from the mediating particle $\eta$. 
The Feynman diagrams for the case $\chi\chi\to Ze\bar e$ are shown in Fig.~\ref{fig:Z_Diagram}.
For the cross-section, we find
in the limit $m_f,v\to 0$
\begin{eqnarray}\label{doubletZff}
\frac{vd\sigma(\chi\chi\to Z f\bar f)}{dE_Z dE_f} & = & \frac{N_c\alpha_{em}f^4 }{2\pi^2 \sin^2(2\theta_W) m_{\rm DM}^4 (2x-x_0^2)^2(1-2y-\mu_f)^2(3-2x-2y+\mu_f)^2}  \nonumber \\
&& {} \times \Big\{ (1+x_0^2/4-x)[x^2-2x(1-y)+2(1-y)^2-x_0^2/2] C_f(x)^2 \nonumber \\
&& {} + x_0^2(1-y-x/2)^2 \big[ (1+x_0^2/4-x)(4C_f(x)+x_0^2) \nonumber \\
&& {} - x_0^2/2 -2(1-y)(1-x-y) \big] \Big\} \;,
\end{eqnarray}
where $ C_f(x) \equiv 1 + \mu_f + (x-x_0^2/2) (g_A^f \pm g_V^f ) -x_0^2/2$, and $g_V^f=t_{3f}-2q_f\sin^2(\theta_W)$ and $g_A^f=t_{3f}$ are the couplings to the $Z$ boson. The plus and minus sign
applies to annihilation into left-handed or right-handed fermions, respectively. For the annihilation into $Z f_L\bar f_L$
the corresponding mediating particle $\eta_f$ has to have the SM gauge quantum numbers of $f_R$, and vice versa.

Similarly, for the annihilation into $W$ bosons, we find
\begin{eqnarray}
\frac{vd\sigma(\chi\chi\to W f_L\bar f_L')}{dE_W dE_f} & = & \frac{N_c\alpha_{em}f^4}{64\pi^2\sin^2(\theta_W) m_{\rm DM}^4 (2x-x_0^2)^2(1-2y-\mu_f)^2} \nonumber \\
&& {} \times \Big\{ 4(1-x)[x^2-2x(1-y)+2(1-y)^2] + 2(1-x-y)x_0^4\nonumber \\
&& {} +x_0^2[5x^2-2x-2+8(1-y)(1-x))] +x_0^6/4 \Big\} \;, \\
\frac{vd\sigma(\chi\chi\to W f_R\bar f_R')}{dE_W dE_f} & = & 2c_W^2 \frac{vd\sigma(\chi\chi\to Z f\bar f)}{dE_Z dE_f} \Big|_{M_Z\mapsto M_W, f^4\mapsto (ff')^2, C_f(x)\to 1+\mu_f-x_0^2/2} \;. \nonumber \\
\end{eqnarray}
The former process can proceed via a mediating particle with the quantum numbers of the right-handed
partners of either $f$ or of $f'$ and incorporates contributions from initial and final state radiation. The latter process is mediated by the doublet $(\eta_f,\eta_f')$ with
quantum numbers of the left-handed fermion doublet $(f,f')$, and involves contributions from initial state radiation and from diagrams where the $W$ is emitted off the internal line. Here we assumed $m_{\eta_f}=m_{\eta_{f'}}$ for simplicity.
Note that annihilation into $W f_R\bar f_R'$ is only possible for quarks in the SM. It can also exist for
leptons if neutrinos are Dirac particles.

In connection to Sommerfeld corrections also annihilations of the charged components of the doublets containing the
dark matter particle are relevant. Here it is important to project out only the contributions where the initial
particles have total spin zero. We find that
\begin{equation}
  \frac{vd\sigma(\chi^+\chi^-\to \gamma f\bar f)|_{S=0}}{dE_\gamma dE_f} = 4s_W^2c_W^2 \frac{vd\sigma(\chi\chi\to Z f\bar f)}{dE_Z dE_f}|_{M_Z\to 0, C_f\to 1+\mu+q_fx} \;.
\end{equation}
For annihilation into right-handed fermions $f_R$, the mediating particle $\eta_f$ needs to have the same
quantum numbers as $f_R$. For annihilation into left-handed fermions $f_L$, a mediating particle
with quantum numbers of the $SU(2)_L$ partner $f_L'$ of $f_L$ is required.

\end{appendix}

\end{document}